\documentclass[12pt, journal,doublecolumn, final]{IEEEtran}
\pdfoutput=1

\usepackage{mymacros}

\newcommand{\xmark}{\ding{55}}%
\newcommand{\cmark}{\ding{51}}%

\newcommand{\Fig}[1]{Fig.~\ref{fig:#1}}
\newcommand{\Tab}[1]{Tab.~\ref{tab:#1}}

\newcommand{\App}[1]{Appendix~\ref{app:#1}}

\newtheorem{proposition}{Proposition}

\newtheorem{corollary}{Corollary}

\DeclareMathOperator*{\argmin}{\arg\!\min}

\acrodef{MNO}{mobile network operator}
\acrodef{PHP}{Poisson hole process}
\acrodef{GPP}{Gauss-Poisson process}
\acrodef{DPP}{determinantal point process}
\acrodef{PPP}{Poisson point process}
\acrodef{MC}{Monte Carlo}
\acrodef{LGCP}{Log-Gaussian Cox process}
\acrodef{SINR}{signal-to-noise and interference ratio}
\acrodef{SIR}{signal-to-interference ratio}
\acrodef{RRH}{Remote Radio Head}
\acrodef{mmWave}{millimetre-wave}
\acrodef{MOCN}{Multi-Operator Core Network}
\acrodef{GWCN}{Gateway Core Network}
\acrodef{Ofcom}{Office for Communications}
\acrodef{3GPP}{3\textsuperscript{rd} Generation Partnership Project}
\acrodef{FCC}{Federal Communications Commission}
\acrodef{pdf}{probability distribution function}
\acrodef{ccdf}{complementary cumulative distribution function}
\newcommand{\yale}[1]{``#1''}

\begin{document}

\title{Radio Access Network and Spectrum Sharing in Mobile Networks: A Stochastic Geometry Perspective}

\author{\IEEEauthorblockN{Jacek Kibi\l{}da, Nicholas J. Kaminski and Luiz A. DaSilva}

\thanks{J. Kibi\l{}da, N. J. Kaminski, and L. A. DaSilva are all with CONNECT, Trinity College, The University of Dublin, Ireland, email:  \{kibildj,dasilval\}@tcd.ie, and mnfork@gmail.com.}% <-this % stops a space
%\thanks{Manuscript received April 19, 2005; revised January 11, 2007.}
\vspace*{-0.5cm}}

\maketitle

\begin{abstract}

Next generation mobile networks will rely ever more heavily on resource sharing. In this article we study the sharing of radio access network and spectrum among mobile operators. We assess the impact of sharing these two types of resources on the performance of spatially distributed mobile networks. We apply stochastic geometry to observe the combined effect of, for example, the level of spatial clustering among the deployed base stations, the shared network size, or the coordination in shared spectrum use on network coverage and expected user data rate. We uncover some complex effects of mobile network resource sharing, which involve non-linearly scaling gains and performance trade-offs related to the sharing scenario or the spatial clustering level.

\end{abstract}

\section{Introduction}

Most studies of mobile network resource sharing focus on techno-economic aspects of sharing, while taking for granted the sharing gains to network coverage and capacity \cite{FrisancoTafertshoferLurinAng_2008,MeddourRasheedGourhant_2011}, relying on an intuition that the more resources we share the better \cite{EC_2012}. This intuition is predicated on over-provisioning of coverage and capacity, and, as such, tells us nothing about the intrinsic characteristics of a shared mobile network or a network sharing scenario. The goal of this work is to build a framework to assess technical trade-offs in radio access network, referred to also as infrastructure, and spectrum sharing in spatially distributed mobile networks.% What we uncover is a complex reality of mobile resource sharing which involves non-linearly scaling gains and performance trade-offs related to the sharing scenario or the spatial clustering level in the radio access network deployment.

\subsection{Related work}

During the recent decade resource sharing in mobile networks has grown in popularity and it is considered now an important cost reduction measure for mobile network operators \cite{MeddourRasheedGourhant_2011}. Indeed, both industry and research communities have already recognized the importance of network sharing in the evolution of mobile networks. The \ac{3GPP} has defined standards for network sharing \cite{3gpp23.251_2014}: the \ac{MOCN}, and the \ac{GWCN}. Furthermore, in 2012 the European Commission issued a report making a case for spectrum and facility sharing in the European market \cite{EC_2012}. Communications regulators have also been testing the feasibility of various spectrum sharing models, as is the case with, for example, the ruling of the \ac{FCC} to open military frequencies in the~\unit[3550-3700]{MHz} band to mobile broadband services\footnote{The ruling can be accessed here: \url{http://transition.fcc.gov/Daily_Releases/Daily_Business/2015/db0421/FCC-15-47A1.pdf}}. Finally, there are numerous commercial examples of mobile network resource sharing.

In the literature to date mobile network resource sharing has been analyzed from the perspective of techno-economic feasibility, a multi-operator resource allocation problem, or a simulation-based performance study. The techno-economic feasibility of sharing was thoroughly analyzed in the works of Markendahl et al. (see, for example, \cite{Markendahl_2011,AhmedYangSungMarkendahl_2015}). Mobile network resource sharing in the form of a~multi-operator resource allocation problem was studied in a number of articles that tackle infrastructure allocation \cite{KibildaDaSilva_2013,CanoCaponeCarelloCesena_2015,AntonopoulosKartsakliBousiaEtAl_2015}, spectrum allocation \cite{HewWhite_2008,JorswieckBadiaFahldieckKaripidisEtAl_2014}, virtual resource sharing \cite{TseliouAdelantadoVerikoukis_2016}, or the interchangeability between infrastructure and spectrum resources \cite{MarottaKaminskiGomes-MiguelezGranvilleEtAl_2015,MacalusoAhmadiGomez-MiguelezEtAl_2016}. Simulation studies of the sharing gains over grid-allocated infrastructure was performed for a range of spectrum and infrastructure sharing scenarios \cite{MiddletonHooliTolliLilleberg_2006,JorswieckBadiaFahldieckKaripidisEtAl_2014,PanchalYatesBuddhikot_2013,LuotoPirinenBennisEtAl_2015}.

In our previous work \cite{KibildaDiFrancescoMalandrinoDaSilva_2015}, we explored, with the help of stochastic geometry, the fundamental resource sharing trade-offs in spatially distributed mobile networks. Application of stochastic geometry to mobile network resource sharing can be traced back to \cite{HuaLiuPanwar_2012}, where the performance of infrastructure sharing in a two-operator \ac{PPP} network is studied, and \cite{LinAndrewsGhosh_2014}, where spectrum sharing between device-to-device networks and macrocell users is considered. At the same time as our work \cite{KibildaDiFrancescoMalandrinoDaSilva_2015}, in \cite{YangSung_2015} stochastic geometry was applied to a study of the interchangeability between spectrum and base station density. Building on our work, stochastic geometry models were applied to study the impact of distance-dependent blockage on the sharing performance \cite{WangSamdanisPerezDiRenzo_2016}, and in \cite{GuptaAndrewsHeath_2015,RebatoMezzavillaRanganZorzi_2016} to make the case for unmanaged spectrum sharing in \ac{mmWave} bands. Spectrum sharing for \ac{mmWave} bands has become a topic in its own right, with different research groups, for example, \cite{GuptaAndrewsHeath_2015,GuptaAlkhateebAndrewsHeath_2016,BoccardiShokri-GhadikolaeiFodorEtAl_2016}, studying different spectrum sharing schemes which exploit features specific to \ac{mmWave} bands, such as pencil-wide antenna beams and distance-dependent shadowing.

\subsection{Contributions}

The subject of this article are the resource sharing trade-offs in spatially distributed multi-operator mobile networks, extending our previous work in \cite{KibildaDiFrancescoMalandrinoDaSilva_2015}. In the original article we studied basic coverage and rate trade-offs for the scenarios of unmanaged \textit{infrastructure sharing}, \textit{spectrum sharing}, and \textit{full sharing} (when both spectrum and radio access infrastructure are shared). We considered two mobile operators whose networks were deployed either \emph{independently}, or with \emph{clustering}, i.e., with positive spatial correlation across base stations belonging to different operators. The clustering case is of particular interest as a model representative of real multi-operator radio access network deployments (see \cite{KibildaGalkinDaSilva_2015}). We also considered an asymptotic case of clustering, i.e., \emph{co-location}, where the nearest base stations of sharing operators are located an arbitrarily small distance apart.

We extend our previous article \cite{KibildaDiFrancescoMalandrinoDaSilva_2015} in three main ways. First, we differentiate between the sharing of spectrum bands experiencing flat or frequency-selective power fading, and provide a new analytical expression to describe the latter case. Second, we investigate the impact of exclusion zone-based coordination in inter-operator spectrum sharing. And, finally, we consider the impact of network density imbalance between the sharing operators on coverage.

We summarize the major contributions and findings of our work as:
\begin{itemize}
\item We provide mathematical expressions for the coverage probability and the average user rate for each of the sharing scenarios, for a shared network created as a union of $|\mathcal{N}|$ independently distributed networks.
\item We show that infrastructure and spectrum sharing cannot be simply substituted for each other, as there exists a trade-off in coverage and rate between the two. Moreover, the combination of the two approaches does not simply produce linearly scaling gains, as the increase in rate is traded for a minor reduction in coverage (when compared to infrastructure sharing performed in isolation).
\item We show that sharing of spectrum bands experiencing frequency-selective fading yields increased coverage over sharing of bands experiencing flat fading. Moreover, when bands are shared under the former case, there exists a cross-over point between the full and infrastructure sharing, which is present only when the channel has many strong multi-path components, e.g., a Rayleigh fading channel.
\item We show that when infrastructure sharing occurs between networks with different densities, it is significantly more beneficial to the user of a smaller network. The opposite is true when only spectrum sharing is considered. 
\item We show that an increase in spatial clustering deteriorates coverage and rate for full and infrastructure sharing scenarios. However, when only spectrum is shared, we observe that there exists a cross-over point, i.e., the stronger the clustering the better coverage at low \ac{SIR} values and the worse coverage at high \ac{SIR} values.
\item We show that even a simple mechanism of coordination in spectrum sharing allows networks to achieve significant improvement in coverage without hampering the rates achievable by their users.
\end{itemize}

\section{System model and methodology}

\subsection{System model}

We model transmitter deployment by a single operator as a point process defined as a random countable set\footnote{Using the random set formalism we implicitly assume that transmitters from more than one operator do not occupy the same location. Yet, this does not preclude transmitters to be located arbitrarily close to each other, which we consider as co-location.} $\Phi_n\subset \mathbb{R}^2$, with elements being random variables $x_i$ and $n\in\mathcal{N}$, with $\mathcal{N}$ denoting the set of mobile operators. Each mobile operator holds a license to an individual spectrum band (each spectrum band is of equal size and does not overlap with any other operator's spectrum band), and, as such, we will often use $n\in\mathcal{N}$ to refer to a spectrum band licensed to operator $n$. We also denote $\Phi=\bigcup_{n\in\mathcal{N}}\Phi_n$ as the set of all transmitter locations of all operators. %Let also $d_{n}(x)=||y_n - x||$ denote the distance from a reference user of operator $n\in\mathcal{N}$ located in $y_n$, which in our system corresponds to the typical point $(0,0)$, to a transmitter in $x\in\Phi_m$ belonging to operator $m\in\mathcal{N}$.
We assume that the power of the signal received by the reference user $y_n$ of operator $n\in\mathcal{N}$ is affected by large- and small-scale fading: pathloss $l(x)$ and power fading $h_x$, where $x$ denotes the location of the serving transmitter. The pathloss function $l:\mathbb{R}^2\rightarrow \mathbb{R}_+$ is of the form $l(x)=||x-y_n||^{-\alpha}$, where $\alpha$ is the pathloss exponent, while the power fading between the transmitter in $x$ and $y_n$ is spatially independent and exponentially distributed with unit mean, i.e., we assume Rayleigh fading with unitary transmit power. These specific channel propagation assumptions are in line with best practice modelling of wireless networks in cellular bands \cite{AndrewsGuptaDhillon_2016}, and allow us to focus our analysis on the fundamental resource sharing trade-offs in spatially distributed networks. In the numerical analysis, we evaluate the impact that power fading has on the system performance by comparing results under Rayleigh fading to that under Nakagami-m fading. Analytical considerations of the impact of shadowing and generalized power fading models on the wireless network performance in cellular bands can be found in \cite{GaliottoEtAl_2014} and \cite{DiRenzoGuidottiCorazza_2013}, respectively. In addition, the impact of blockage on the performance of shared mobile networks has been recently considered in~\cite{WangSamdanisPerezDiRenzo_2016}.

We assume that the reference user will associate with the closest transmitter of operator $n$, i.e., $\argmin_{x\in\Phi_n} ||x-y_n||$, which implies that, when infrastructure sharing is in place, the reference user of operator $n$ will associate with any network of any infrastructure sharing operator, i.e., the association rule becomes $\argmin_{x\in\Phi} ||x-y_n||$. Note that, since the point processes we consider in the following are stationary, we assume that the reference user is always located in the origin. Moreover, our model applies to the downlink only and presupposes a full buffer transmission, i.e., each transmitter continuously operates in the selected band(s).

\subsection{Performance metrics}

In our work we assess the efficiency of resource sharing utilizing two performance metrics: coverage probability and average user rate. The former represents the \ac{ccdf} of the \ac{SINR} and is defined as:
\begin{equation}
p(\theta) = \mathbb{P} \big(\mathrm{SINR} > \theta\big),
\label{eq:pc}
\end{equation}
where $\theta$ is the reception threshold at the physical layer (given a linear receiver with interference treated as noise). Assuming exponential fading, the expression in Eq.~(1) can be transformed into \cite{AndrewsBaccelliGanti_2011}:
\begin{equation}
p(\theta) = \int_{0^+}^\infty \exp(-sW) \mathcal{L}_{I}(s) f_R(r)dr,
\label{eq:cp_general}
\end{equation}
where $s = \theta/\tilde{l}(r)$, $\tilde{l}(r) \equiv l(x)$ with $x\in\Phi$, and $\mathcal{L}_{I}(s)$ is the Laplace transform of interference $I$; $f_{R}(r)$ is the distribution of distance to the serving base station in $x$. For a \ac{PPP} network $\Phi$ with intensity $\lambda$ the Laplace transform of interference is known \cite{AndrewsBaccelliGanti_2011}, and it can be expressed as:
\begin{equation}
\mathcal{L}_{I}(\theta r^\alpha) = \exp(-\pi r^2 \lambda \mathfrak{Z}(\theta,\alpha)),
\label{eq:laplace_general}
\end{equation}
where $\alpha > 2$, $r$ denotes the distance to the nearest base station, and $\mathfrak{Z}(\theta,\alpha) = \frac{2\theta}{\alpha - 2}\mbox{$_2$F$_1$}(1, 1-2/\alpha;\,2-2/\alpha;\,-\theta)$, where $\mbox{$_2$F$_1$}(a, b;\,c;\,z)$ is the Gauss hypergeometric function.

The average user rate can be defined as the expected rate of a user when adaptive modulation and coding is set so that the Shannon bound\footnote{Real-world mobile systems do not achieve this bound but to account for this fact would simply require that we rescale our results.} is achieved for the instantaneous \ac{SINR} of that user:
\begin{equation}
d = \mathbb{E} \left[ b\log_2\big(1 + \mathrm{SINR}\big) \right],
\label{eq:dr_def}
\end{equation}
where $b$ denotes the spectrum bandwidth. This average user rate may then be expressed in terms of the coverage probability \cite{AndrewsBaccelliGanti_2011}:
\begin{equation}
d = \mathbb{E}_\rho \left[ p\big(2^{\frac{\rho}{b}}-1\big)\right].
\label{eq:adr_general}
\end{equation}
In the numerical results, we also consider user rate at the $5^{\text{th}}$, $50^{\text{th}}$ (median), and $95^{\text{th}}$ percentile. The intention is to inspect how positive or negative effects of sharing are distributed across different users.

\subsection{Scenarios}

\begin{figure*}[tb!]
\centering
\subfigure[Exclusive use\label{fig:dyspan_no_sharing}]{
 \includegraphics[width=0.4\textwidth]{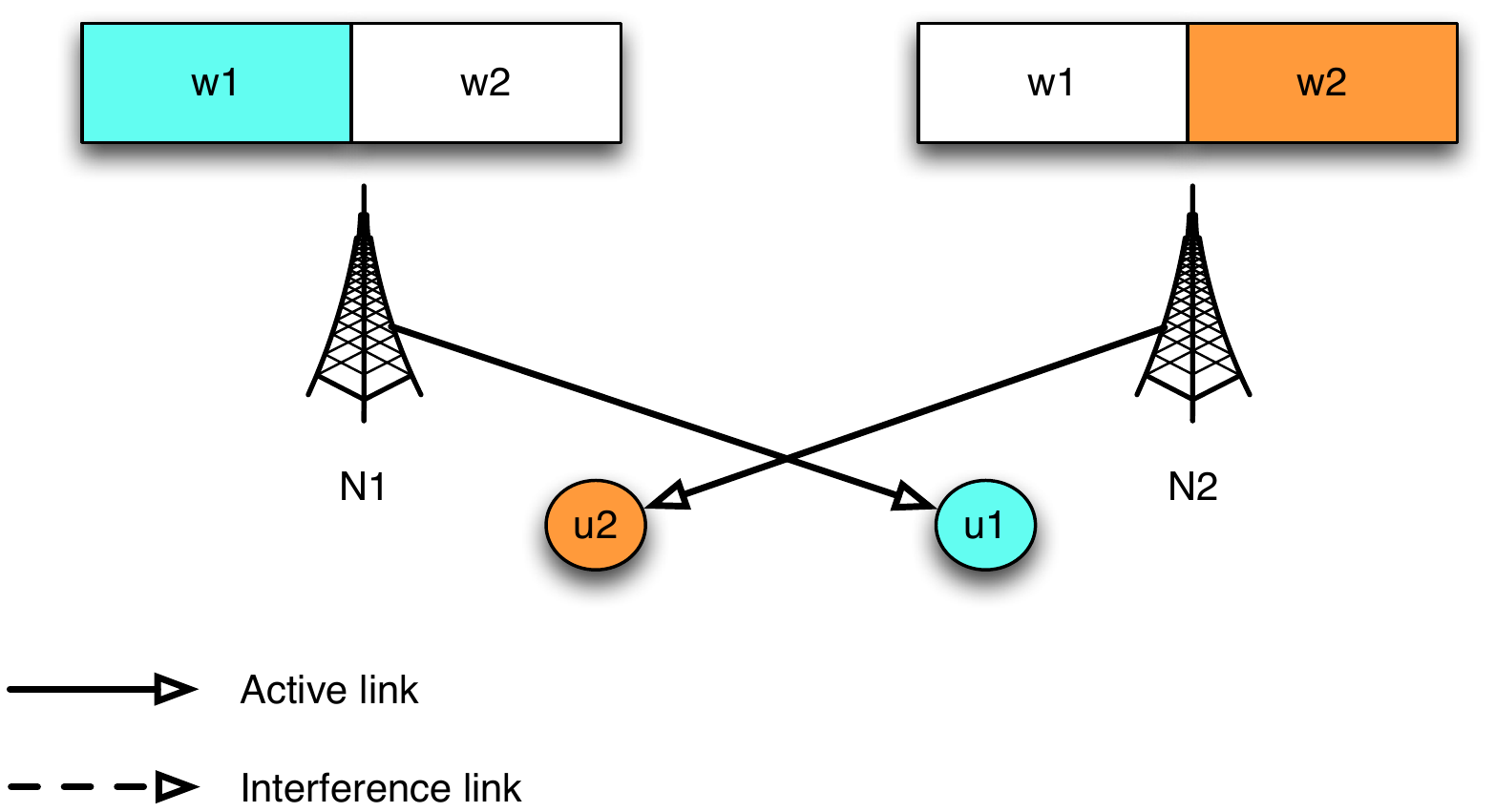}
}
\subfigure[Infrastructure sharing\label{fig:dyspan_infrastructure_sharing}]{
 \includegraphics[width=0.4\textwidth]{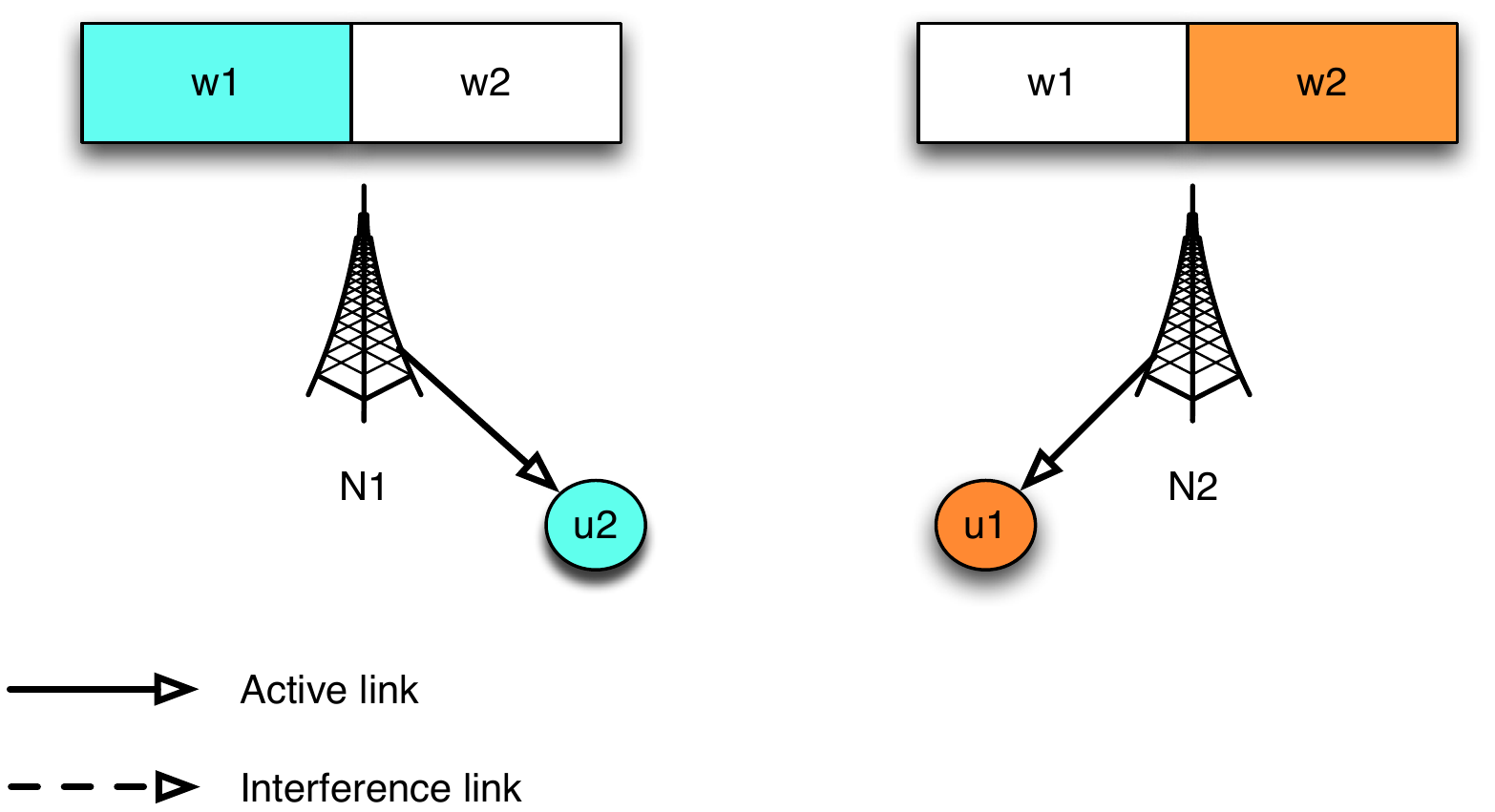}
}

\subfigure[Spectrum sharing\label{fig:dyspan_spectrum_sharing}]{
 \includegraphics[width=0.4\textwidth]{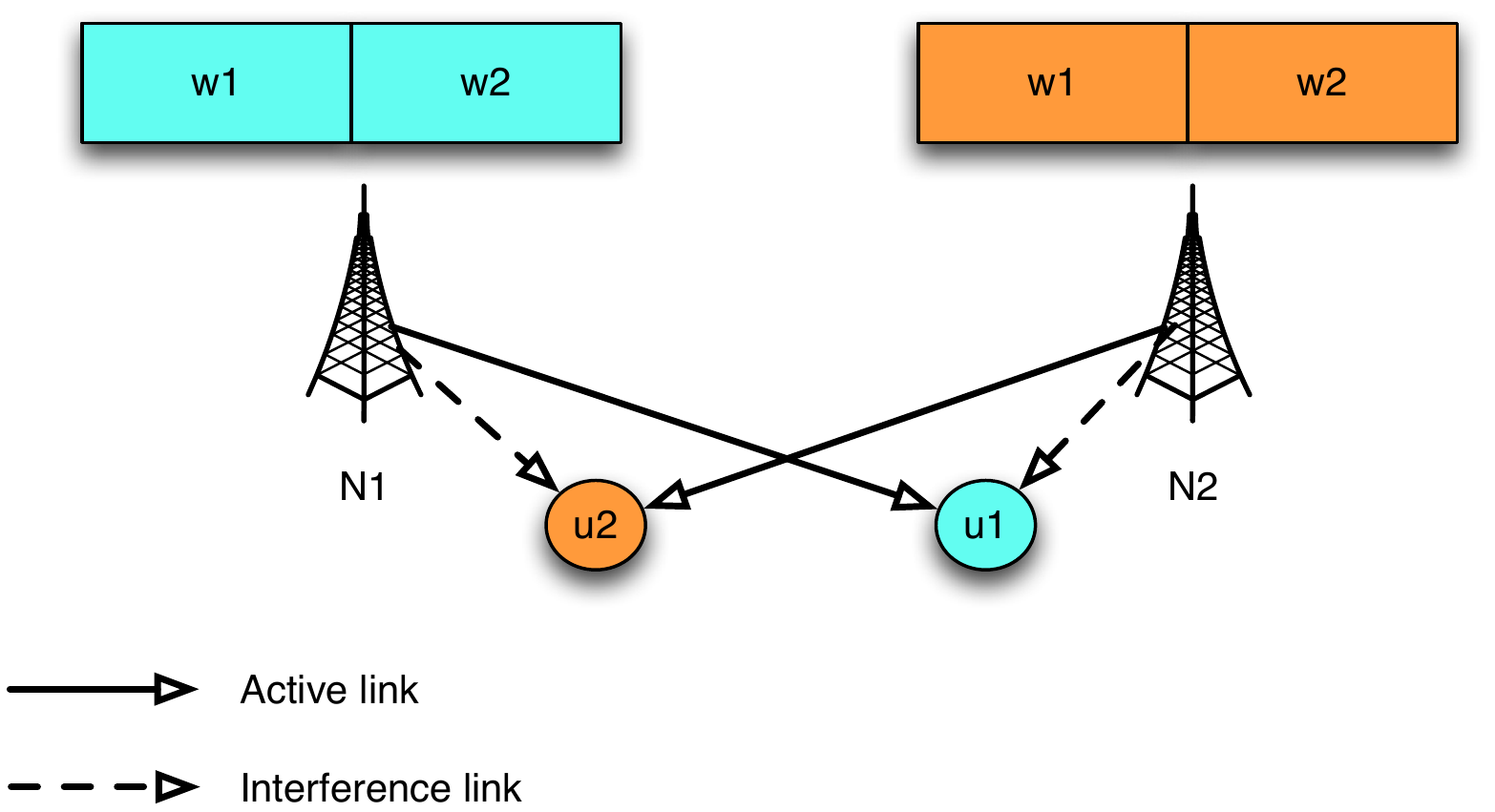}
}
\subfigure[Full sharing\label{fig:dyspan_all_sharing}]{
 \includegraphics[width=0.4\textwidth]{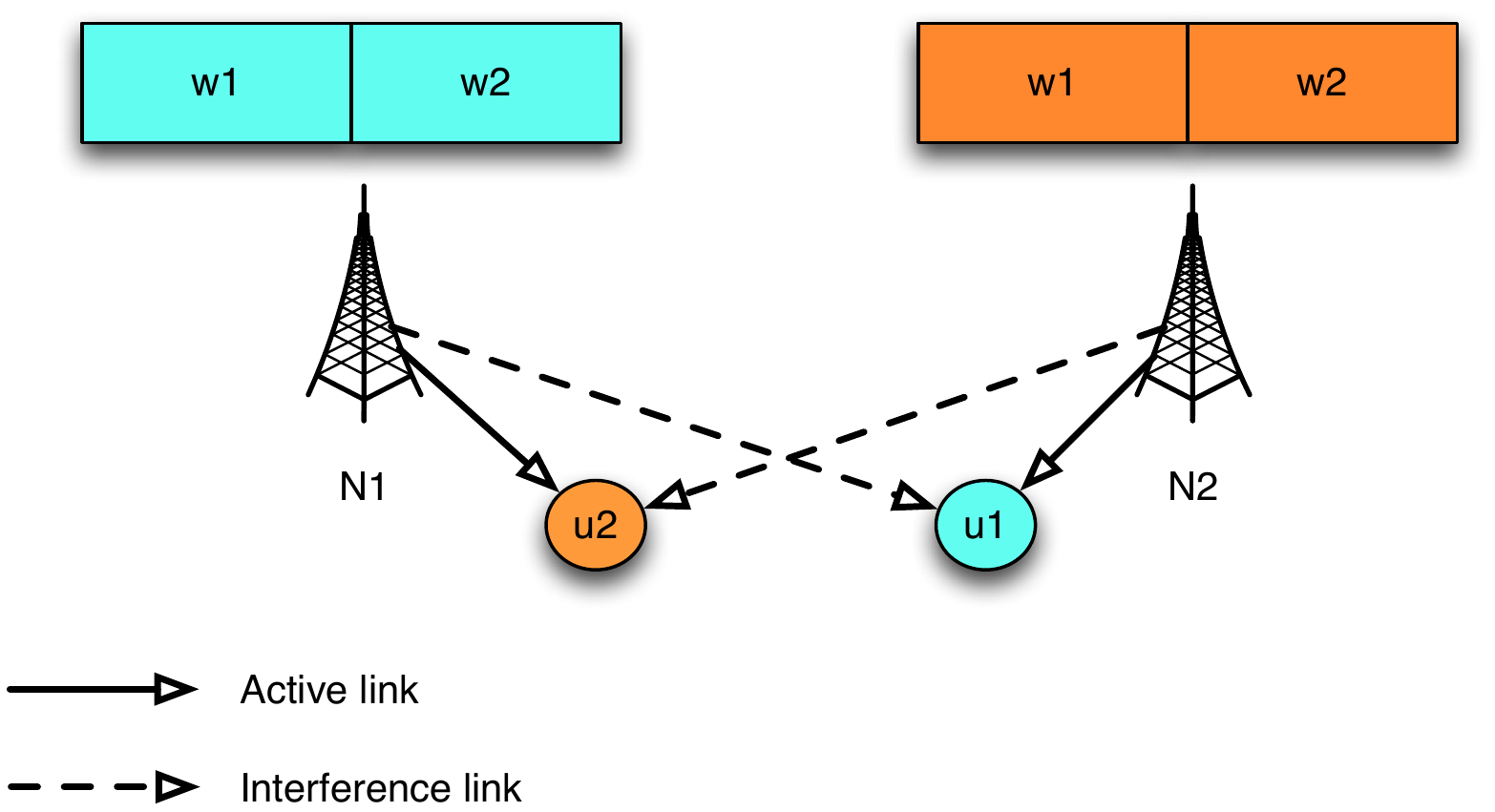}
}
\caption{
Two operators (1 and 2) with subscribers (u1, u2): (a) exclusively using their spectrum (w1, w2), (b) allowing for access to each other's infrastructure (N1, N2), (c) allowing for shared use of spectrum, and (d) allowing for both shared use of spectrum and infrastructure.
}
\end{figure*}

The conventional scenario of a commercial mobile market consists of competing operators, each of which owns the network infrastructure and holds a license to one spectrum band, which also implies that subscribers of one operator exclusively\footnote{Leaving international roaming out of the picture.} utilize the infrastructure and spectrum of their operator as depicted in \Fig{dyspan_no_sharing}. In this paper, we study inter-operator resource sharing and, thus, we consider a number of network sharing scenarios. At the highest level of abstraction, we define three network sharing scenarios of \textit{infrastructure sharing}, \textit{spectrum sharing}, and \textit{full sharing}, where both radio access network and spectrum are shared. 

In the \emph{infrastructure sharing} scenario, we assume that operators pool their radio access infrastructure without pooling their operational frequencies (see \Fig{dyspan_infrastructure_sharing}). Effectively, interference to the desired signal is identical to the single-operator case, i.e., it comes from $\Phi_n$ where $n$ is the operator to which the serving transmitter belongs. In the \emph{spectrum sharing} scenario, we assume that operators allow access to their spectrum bands from outside of their network, without sharing their infrastructure (see \Fig{dyspan_spectrum_sharing}). Effectively, interference to the desired signal of a user of operator $n$ in spectrum band $k\in\mathcal{N}$ comes from all transmitters of operator $k$ and the transmitters of all other operators. In the \emph{full sharing} scenario, operators pool their radio access infrastructure and allow for shared use of spectrum. Spectrum pooling results in increased interference, as more transmitters may now operate in the same frequency band and users may connect to any of the transmitters of the sharing operators (see \Fig{dyspan_all_sharing}).

For the scenarios where spectrum is shared, i.e., \emph{spectrum} and \emph{full sharing}, we consider two cases of operators aggregating bands that experience: flat fading, and frequency-selective fading. All these scenarios include worst-case interference, i.e., \emph{uncoordinated use of shared spectrum} among operators, hence we also study the scenario of \emph{coordinated shared spectrum} use. In order to coordinate shared spectrum use among transmitters of the sharing operators, we apply exclusion zones, i.e., discs of arbitrary size centred at the locations of transmitters. Formally, a spectrum sharing exclusion zone for a spectrum band $n$ is a disc of radius $\mathbb{R}^2s\in\mathbb{R}^+\cup\{0\}$ centred at the location of a transmitter $y\in\Phi_n$ which prohibits transmitters of any operator other than $n$ located within that disc from utilizing spectrum band $n$. For the sake of simplicity in the resulting expressions, we assume that $\mathbb{R}^2s$ is the same for all operators, yet a more general case requires simply an additional index. Now, some transmitter $x$ utilizes spectrum band $n\in\mathcal{N}$ according to the following policy:
\begin{equation}
v^n_x = \begin{cases} 1, & \mbox{if } x\in\Phi_n, \\ 1, & \mbox{if } x\in\Phi\setminus\Phi_n \vee ||x - y|| > \mathbb{R}^2s, \forall y\in\Phi_n, \\
0, & \mbox{otherwise}, \end{cases}
\end{equation}
where $v^n_x$ denotes the spectrum use policy for transmitter $x$ for spectrum band $n$.% The policy implies that only two types of transmitters may operate in spectrum band $n$: all the transmitters of operator $n$, i.e., $\Phi_n$, and transmitters of operators other than $n$ which are at least an $\mathbb{R}^2s$ distance apart from any transmitter of operator $n$.

In \Tab{estimated_params} we summarize all the scenarios; a check mark should be interpreted as an indication that a specific feature is considered under a given scenario.

\begin{table*}[tb]
%\renewcommand{\arraystretch}{1.3}
% if using array.sty, it might be a good idea to tweak the value of
% \extrarowheight as needed to properly center the text within the cells
\caption{Summary of features included in the studied scenarios}
\label{tab:estimated_params}
\centering
%\multirow{2}{*}{Location type}
%\multicolumn{3}{c}{Model parameters}
\begin{tabular}{ccccc}
\hline
scenario & infrastructure & spectrum & coordinated use of spectrum & flat/frequency-selective fading\\
\hline
\emph{baseline} & \xmark & \xmark & \text{NA} & NA \\
\emph{infrastructure sharing} & \cmark & \xmark & NA & NA \\
\emph{spectrum sharing} & \xmark & \cmark & \cmark & \cmark \\
\emph{full sharing} & \cmark & \cmark & \cmark & \cmark \\
\hline
\end{tabular}
\end{table*}

\section{Analytical results}
\label{sec:analytical_results}

In this section, we apply stochastic geometry to derive analytical expressions for the coverage probability\footnote{We express the average user rate in terms of the coverage probability, according to the relationship in Eq.~(5).} for the scenarios of infrastructure, spectrum and full sharing between multiple mobile operators. We assume each mobile operator owns radio access infrastructure distributed according to a uniform \ac{PPP} with intensity $\lambda_n$, and has a license to use one spectrum band of bandwidth $b$. Whenever spectrum is shared, we differentiate between the sharing of spectrum bands that experience flat and frequency-selective fading. We also assume constant reception thresholds and constant noise levels across all shared spectrum bands, and, finally, we present all our results from the perspective of a reference user of operator $n$, always providing a general form and, whenever possible, an easy-to-interpret closed-form that emerges when an interference-limited regime is considered and $\alpha=4$.

\subsection{Baseline scenario}

Expressions for the coverage probability and the average user rate for this scenario exist in the literature (see \cite{AndrewsBaccelliGanti_2011}) and are given in Eq.~(2) and Eq.~(5), respectively.

\subsection{Infrastructure sharing scenario}

In this scenario users connect to the closest transmitter belonging to the network of any of the $|\mathcal{N}|$ operators. The \ac{SINR} for this scenario for a subscriber of operator $n$ takes the following form:
\begin{equation}
\mathrm{SINR}_k = \frac{h_{x}l (x)}{W + \sum_{y\in\Phi_n\setminus \{x\}}h_y l(y)},
\label{eq:sinr_roaming_polar}
\end{equation}
where $W$ is the noise power, $x\in\Phi_n$, and $k\in\mathcal{N}$. Note that $\mathrm{SINR}_k=0$ for any $k\neq n$.

\begin{proposition}
The coverage probability of a reference user of operator $n$ in the infrastructure sharing scenario can be expressed as:
\begin{align}
p_n(\theta) & = \pi \sum_{i\in\mathcal{N}}\lambda_i\int_{0}^\infty \exp(-\theta r^{\alpha/2} W) \\\nonumber
& \exp\Big(-\pi r \Big( \sum_{j\in\mathcal{N}}\lambda_j + \lambda_i\mathfrak{Z}(\theta,\alpha) \Big) \Big) dr,
\label{eq:cp_ppp_roam_exp_noise}
\end{align}
where $\mathfrak{Z}(\theta,\alpha) = \frac{2\theta}{\alpha - 2}\mbox{$_2$F$_1$}(1, 1-2/\alpha;\,2-2/\alpha;\,-\theta)$.% The rate probability of a reference user of operator $n$ can be found as in \Eq{adr_general}.
%\begin{equation}
%\tau_n = \mathbb{E}_\gamma \bigg[p_n\Big(\exp(\gamma/b_n) - 1\Big)\bigg].
%\end{equation}

\label{prop:roaming}
\end{proposition}
\begin{IEEEproof}

The proof is provided in \App{infrastructure_sharing}.

\end{IEEEproof}

\begin{corollary}
When no noise and $\alpha=4$ is assumed, the coverage probability of a user of operator $n$ is given as:
\begin{equation}
p_n(\theta) = \sum_{i\in\mathcal{N}}\frac{1}{\sum_{j\in\mathcal{N}}\frac{\lambda_j}{\lambda_i} + \sqrt{\theta}\arctan(\sqrt{\theta})}.
\label{eq:cp_ppp_roam_nonoise}
\end{equation}
\label{cor:cp_ppp_roam_nonoise}
\end{corollary}

%Assuming no noise and $\alpha = 4$, we can calculate the integral in \Eq{cp_ppp_roam_exp_noise}. In addition, $\mathfrak{Z}(\theta,4)$ simplifies to $\sqrt{\theta}\arctan(\sqrt{\theta})$.

\subsection{Spectrum sharing scenario}

The \ac{SINR} for this scenario takes the following form:
\begin{equation}
\mathrm{SINR}_k = \frac{h_{x}l(x)}{\eta W + \sum_{y\in\Phi\setminus \{x\}}h_y l (y)},
\label{eq:sinr_spectrum_polar}
\end{equation}
where $x\in\Phi_n$, $k\in\mathcal{N}$, and $\eta$ is a scalar that accounts for the transmit power splitting across the shared bands, e.g., when $k$ is the only band used by a transmitter $\eta=1$, while if all spectrum bands are used $\eta=|\mathcal{N}|$. %, while for the best channel selection $\eta=1$, similarly to scenarios without spectrum sharing. 
Now, finding the coverage probability for the spectrum sharing scenario will be equivalent to finding the probability:
\begin{equation}
p_n(\theta) = \mathbb{P} \bigg(\bigcap_{k\in\mathcal{N}} \mathrm{SINR}_k > \theta, \eta=|\mathcal{N}| \bigg).
\label{eq:pc_n_general_spectrum_sharing}
\end{equation}
At the same time, the average user rate can then be expressed as:
\begin{equation}
d_n = \mathbb{E} \left[ b\sum_{k\in\mathcal{N}}\log_2\big(1 + \mathrm{SINR}_k\big), \eta=|\mathcal{N}| \right],
\label{eq:adr_general_spectrum_sharing}
\end{equation}
which can be simplified to:
\begin{equation}
d_n = |\mathcal{N}|b\mathbb{E} \left[\log_2\big(1 + \mathrm{SINR}_k\big), \eta=|\mathcal{N}|\right],
\end{equation}
and subsequently to:
\begin{equation}
d_n = \mathbb{E}_\rho \left[ p_n\big(2^{\frac{\rho}{|\mathcal{N}|b}}-1\big), \eta=|\mathcal{N}| \right].
\label{eq:adr_pc_spectrum_sharing}
\end{equation}

\noindent
{\bf Sharing spectrum bands under flat fading}

Thanks to the flat fading assumption, the expression in Eq.~(11) can be simplified to:
\begin{equation}
p_n(\theta) = \mathbb{P} \bigg(\mathrm{SINR}_k > \theta, \eta=|\mathcal{N}| \bigg),
\end{equation}
where $\mathrm{SINR}_k$ is described in Eq.~(10).

\begin{proposition}
The coverage probability of a user belonging to operator $n$ in the spectrum sharing scenario under flat fading can be expressed as:
\begin{align}
& p_n (\theta) = \pi \lambda_n\int_{0}^\infty \exp(-\theta r^{\alpha/2} |\mathcal{N}| W) \exp\Big(-\pi r\nonumber\\
& \Big((1 + \mathfrak{Z}(\theta,\alpha))\lambda_n + \mathfrak{Z}_0(\theta,\alpha)\sum_{j\in\mathcal{N}\setminus\{n\}}\lambda_j \Big) \Big) dr.
\label{eq:cp_ppp_spec_exp_nonoise_a4}
\end{align}
\end{proposition}

\begin{IEEEproof}
The proof is provided in \App{spectrum_sharing}.
\end{IEEEproof}

\begin{corollary}
When no noise and $\alpha=4$ is assumed, the coverage probability of a user of operator $n$ is given as:
\begin{equation}
p_n (\theta) = \frac{1}{1 + \sqrt{\theta} \Big(\arctan(\sqrt{\theta}) + \frac{\pi}{2}\sum_{j\in\mathcal{N}\setminus\{n\}}\frac{\lambda_j}{\lambda_n}\Big)}.
\label{eq:cp_ppp_spec_adj_nonoise}
\end{equation}
\label{cor:cp_ppp_spec_adj_nonoise}
\end{corollary}

The average user rate can be obtained by averaging over the obtained coverage probability expression following Eq.~(14).

\noindent
{\bf Sharing spectrum bands under frequency-selective fading}

When shared spectrum bands experience frequency-selective fading, we get the following result:

\begin{proposition}
The coverage probability of a user belonging to operator $n$ from sharing of spectrum bands under frequency-selective fading can be expressed as:
\begin{align}
&p_n (\theta) = \pi \lambda_n \sum_{k\in\mathcal{N}} \binom{|\mathcal{N}|}{k} (-1)^{k+1} \int_{0}^\infty \exp\Big(-\theta r^{\alpha/2} kW\Big)\nonumber\\ 
& \exp\Bigg(-\pi r \bigg( \lambda_n + \lambda_n   \sum^{k}_{l=1} \binom{k}{l} (-1)^{l+1} \mathfrak{Z}(\theta,\alpha,l) + \nonumber\\ &\sum_{j\in\mathcal{N}\setminus\{n\}}\lambda_j\sum^{k}_{l=1} \binom{k}{l} (-1)^{l+1} \mathfrak{Z}_0(\theta,\alpha,l) \bigg) \Bigg) dr.
\label{eq:cp_ppp_spec_nadj_exp_nonoise_a4_div}
\end{align}
\end{proposition}

\begin{IEEEproof}
The proof is provided in \App{spectrum_sharing_diversity}.
\end{IEEEproof}

\begin{corollary}
When no noise and $\alpha=4$ is assumed, the coverage probability of a user of operator $n$ is given as:
\begin{align}
& p_n (\theta) = \nonumber\\
&\frac{\sum_{k\in\mathcal{N}} \binom{|\mathcal{N}|}{k} (-1)^{k+1}}{1 + \sum^{k}_{l=1} \binom{k}{l} (-1)^{l+1} \bigg(\mathfrak{Z}(\theta,4,l) + \sum_{j\in\mathcal{N}\setminus\{n\}}\frac{\lambda_j}{\lambda_n}\frac{\sqrt{\theta}}{2\sqrt{\pi l}}\bigg) },
\label{eq:cp_ppp_spec_nadj_nonoise}
\end{align}
where $\mathfrak{Z}_0(\theta,4,l) \approx \frac{\sqrt{\theta}}{2\sqrt{\pi l}}$.
\label{cor:cp_ppp_spec_nadj_nonoise}
\end{corollary}

\subsection{Full sharing scenario}

The \ac{SINR} model is essentially the same as in the spectrum sharing scenario, with the only difference being that the tagged transmitter comes from the pool of all infrastructure sharing operators.

\noindent
{\bf Sharing spectrum bands under flat fading}

\begin{proposition}
The coverage probability of a user of operator $n$ in the full sharing scenario, when shared spectrum bands experience flat fading, can be expressed as:
\begin{align}
&p_{n} (\theta) = \pi \Big(\sum_{i\in\mathcal{N}}\lambda_i\Big)\int_{0}^\infty \exp(-\theta r^{\alpha/2} |\mathcal{N}| W)\nonumber\\
&\exp\Big(-\pi r \Big(1 + \mathfrak{Z}(\theta,\alpha) \Big)\sum_{j\in\mathcal{N}}\lambda_j \Big) dr.
\label{eq:cp_ppp_full_exp_nonoise_a4}
\end{align}
\label{prop:full_sharing}
\end{proposition}

\begin{IEEEproof}
Let us first recall the following result: the superposition of independent \acp{PPP} is also a \ac{PPP}, with intensity equal to the sum of the intensities of the component processes \cite{BaccelliBlaszczyszyn_2009}[Proposition 1.3.3]. In the case of the full sharing scenario, the users connect to the closest transmitter of any of the networks of the sharing operators and suffer interference from all the other transmitters. Formally a fully shared network is a \ac{PPP} $\Phi = \bigcup_{i\in\mathcal{N}} \Phi_i$ with intensity $\lambda = \sum_{i\in\mathcal{N}}\lambda_i$. Therefore its coverage probability corresponds to that of the baseline case (single-operator and exclusive use of resources), yet with intensity $\lambda$.
\end{IEEEproof}

When no noise and $\alpha=4$ is assumed, the coverage probability of a user of operator $n$ simplifies to that of a single-operator \ac{PPP} network, expressed in Eq.~(2). The average user rate can be calculated by substituting the obtained coverage probability into Eq.~(14). 

\noindent
{\bf Sharing spectrum bands under frequency-selective fading}

\begin{proposition}
The coverage probability of a user belonging to operator $n$ in full sharing scenario, when shared spectrum bands experience frequency-selective fading, can be expressed as:
\begin{align}
&p_n (\theta) = \pi \sum_{i\in\mathcal{N}}\lambda_i \sum_{k\in\mathcal{N}} \binom{|\mathcal{N}|}{k} (-1)^{k+1} \nonumber\\
& \int_{0}^\infty \exp\Big(-\theta r^{\alpha/2} kW\Big)\exp\Bigg(-\pi r \nonumber\\
& \sum_{j\in\mathcal{N}}\lambda_j\bigg( 1 + \sum^{k}_{l=1} \binom{k}{l} (-1)^{l+1} \mathfrak{Z}(\theta,\alpha,l)\bigg) \Bigg) dr.
\label{eq:cp_ppp_full_nadj_exp_nonoise_a4_div}
\end{align}
\end{proposition}

\begin{IEEEproof}
The derivation of this result mirrors that provided in \App{spectrum_sharing_diversity}, with a modification to the set of interferers.
\end{IEEEproof}

\begin{corollary}
When no noise and $\alpha=4$ is assumed, the coverage probability of a user of operator $n$ is given as:
\begin{equation}
p_n (\theta) = \sum_{k\in\mathcal{N}} \frac{\binom{|\mathcal{N}|}{k} (-1)^{k+1}}{1 + \sum^{k}_{l=1} \binom{k}{l} (-1)^{l+1} \mathfrak{Z}(\theta,4,l)}.
\label{eq:cp_ppp_full_nadj_exp_nonoise_a4_div}
\end{equation}
\label{cor:cp_ppp_full_nadj_exp_nonoise_a4_div}
\end{corollary}

\section{Numerical results and discussion}
\label{sec:results}

Having derived analytical results for the case of independently distributed networks of $|\mathcal{N}|$ operators, we now focus our analysis of resource sharing trade-offs on the case $|\mathcal{N}|=2$. This allows us to model clustered multi-operator deployments using the \ac{GPP} \cite{Haenggi_2013}, which simplifies analysis of inter-operator clustering\footnote{Extension to analysis of shared clustered network when $|\mathcal{N}|>2$ would require that we utilize other than a \ac{GPP} point process model, e.g., one of the Poisson clustered processes \cite{Haenggi_2013}.}. In the \ac{GPP}, clusters of points are distributed according to a \ac{PPP} with intensity $\lambda$. Each cluster consists of either one or two points with probability $1-p$ and $p$, respectively, with one of the points being located in the center of the cluster while the other being uniformly distributed on a circle of radius $u$ surrounding the cluster center. Despite this relatively simple definition, best-known analytical expression for the performance of a \ac{GPP}-distributed network involves integration over the distance to the interfering transmitters, which may be cumbersome to evaluate numerically\footnote{Bounds on the Laplace transform of the \ac{GPP} can be found in \cite{GuoZhongHaenggiZhang_2014}.}. Therefore we evaluate the network sharing performance in a \ac{GPP} distributed multi-operator network using \ac{MC} simulations. It is worth noting that, in both the \ac{PPP} and \ac{GPP} cases, the underlying process of distributing transmitters for a single-operator deployment is the \ac{PPP}, which according to spatial statistical studies to date (see \cite{GuoHaenggi_2013,KibildaGalkinDaSilva_2015}) may be considered a reasonable approximation of a real-world single-operator single-technology deployment.

When presenting our results, we focus on the fundamental trade-offs in wireless network resource sharing, hence, we use the single-slope pathloss model with the pathloss exponent $\alpha=4$ (note that our analytical results are given for any $\alpha>2$) and Rayleigh power fading, which provide a reasonable set of assumptions for cellular network performance analysis \cite{AndrewsGuptaDhillon_2016}. We validate the Rayleigh fading assumption by comparing the performance results from our model to those from a model with Nakagami--m fading. The impact of more general propagation models on the performance of a wireless system was investigated in, for example, \cite{DiRenzoGuidottiCorazza_2013,GaliottoEtAl_2014}. In addition, we assume no noise, i.e., we use the \ac{SIR}, despite the noise power at the receiver being dependent on the number of aggregated spectrum bands (see, for example, Eq.~(16) or Eq.~(18)). The rationale is that most cellular systems are interference-limited -- a rationale that is also followed by the seminal paper on the application of stochastic geometry to cellular networks (see \cite{AndrewsBaccelliGanti_2011}).

\begin{figure*}[tb!]
\centering
\subfigure[Shared bands under flat fading \label{fig:cp_validation_fig1}]{
 \includegraphics[width=0.31\textwidth]{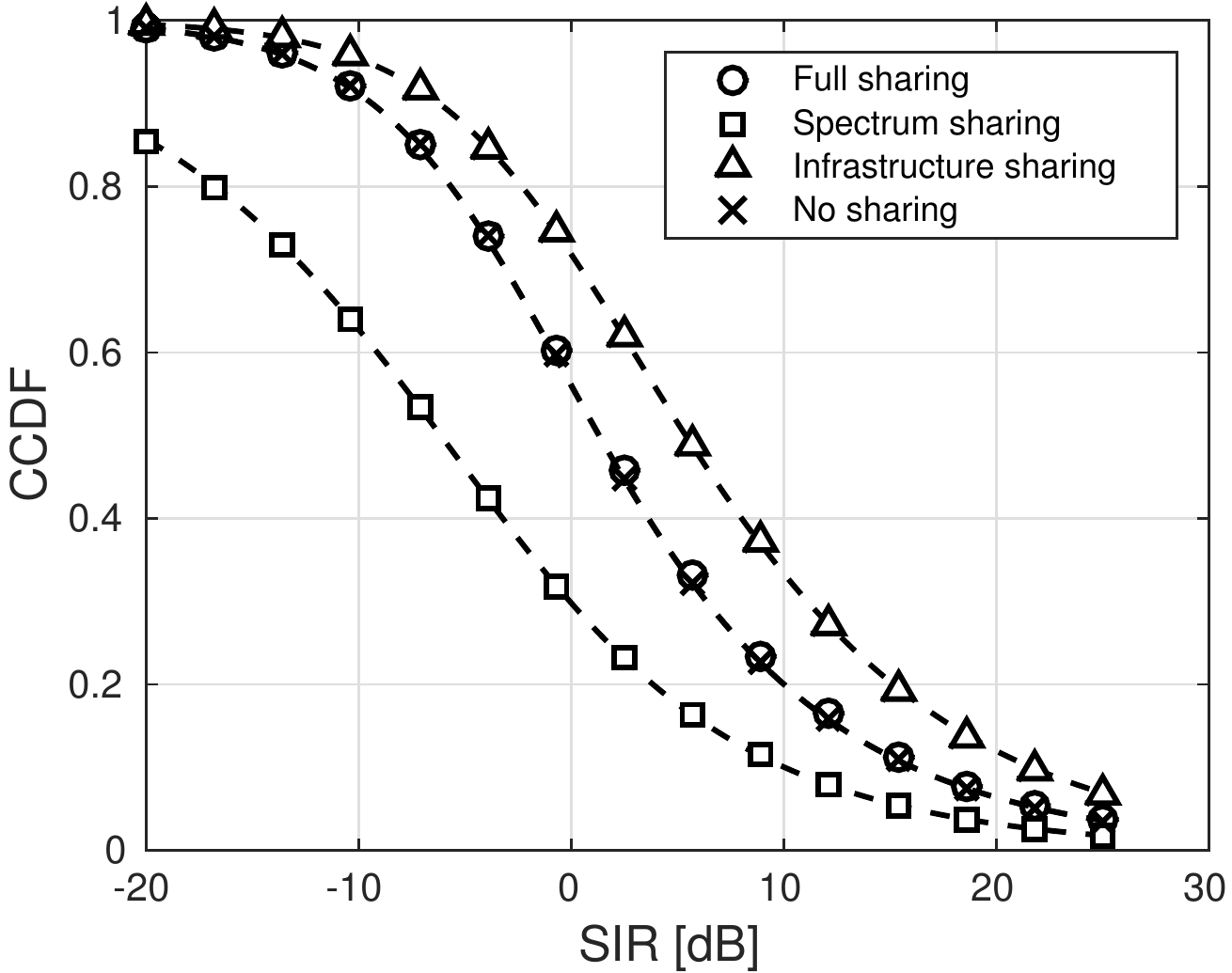}
}
\subfigure[Shared bands under frequency-selective fading \label{fig:cp_validation_fig1_uncorr}]{
 \includegraphics[width=0.3325\textwidth]{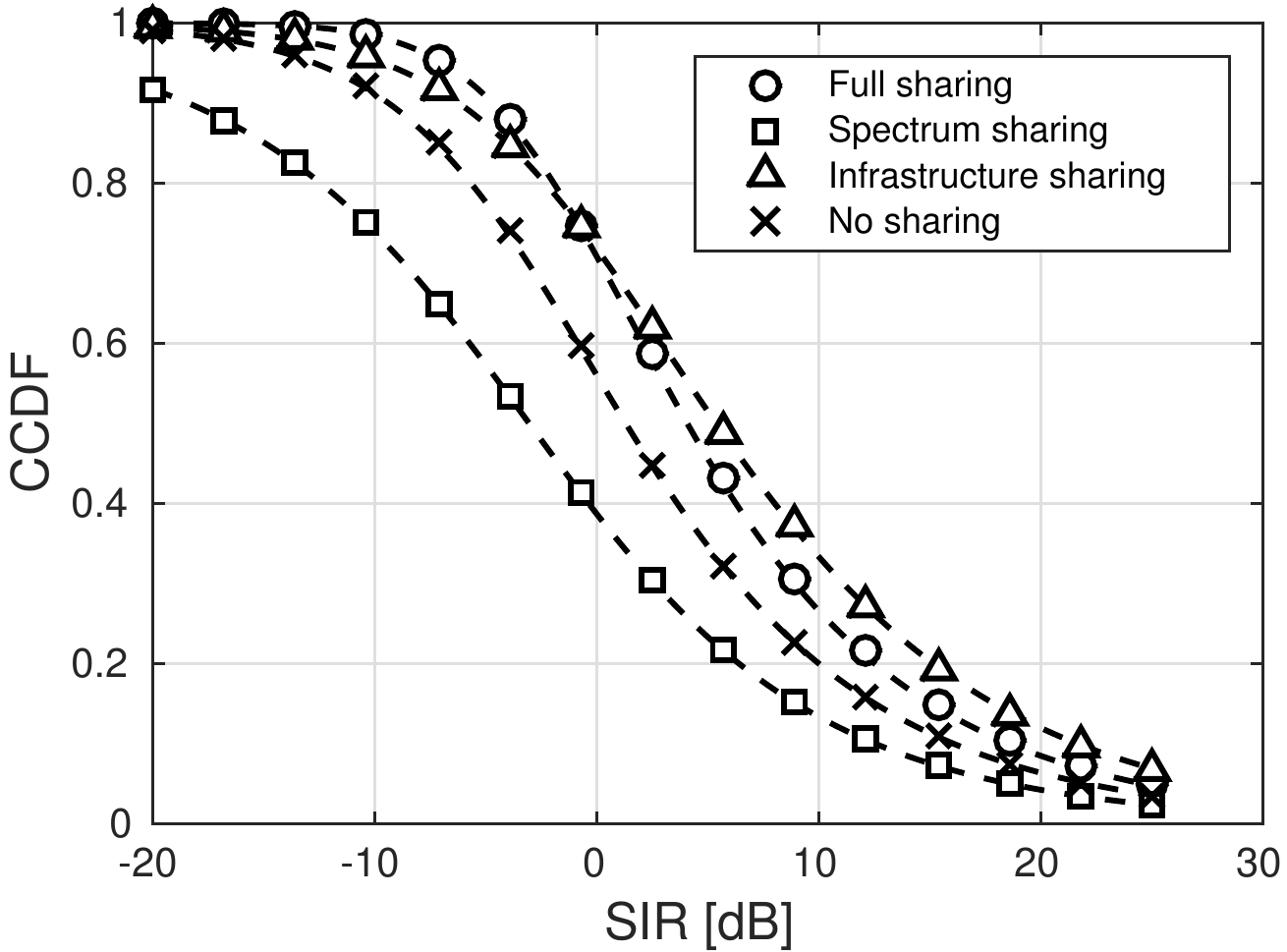}
}
\caption{Coverage probability for different sharing scenarios; markers represent the numerical evaluations, while the dashed lines represent the closed-form expressions. Note that \yale{No sharing} and \yale{Infrastructure sharing} curves remain unchanged from (a) to (b), as only the performance of scenarios with spectrum sharing is affected by flat versus frequency-selective fading assumption across shared bands.}
\end{figure*}

We divide our numerical analysis into five parts. First, we look into the fundamental performance impact of sharing based on our scenarios of infrastructure, spectrum and full network sharing, for the case of independently distributed infrastructure. Then, we look at the impact of network-related parameters, such as the number of sharing operators and the network density, on the sharing performance. Moreover, we validate the impact that our small-scale fading model has on the sharing gains. Subsequently, we relax the assumption of infrastructure independence and analyze the impact of spatial clustering between the infrastructure of different mobile operators. Finally, we inspect the worst-case interference assumption for the unmanaged spectrum sharing case and look at the performance improvement that results from coordination in shared spectrum use. We cross-validate our \ac{MC} simulation results with the derived closed-form expressions (see \Fig{cp_validation_fig1} and \Fig{cp_validation_fig1_uncorr}).

\subsection{Fundamental performance impact of inter-operator resource sharing}

The first thing we analyze is the comparative coverage result for various sharing scenarios, when networks of sharing operators are evenly sized. In \Fig{cp_validation_fig1} we immediately see that infrastructure sharing provides superior coverage over the no sharing and spectrum sharing. This comes from the fact that infrastructure sharing increases the number of transmitters that a user can attach to without affecting the interference in the network. Spectrum sharing significantly increases interference, as there may be active transmitters of another operator located arbitrarily close to the user; this results in the lowest coverage across the four sharing scenarios. When the shared spectrum bands experience frequency-selective fading (see \Fig{cp_validation_fig1_uncorr}), we see the coverage performance in each spectrum sharing scenario being significantly improved, due to diversity gains, with full sharing faring very closely to infrastructure sharing. Interestingly, there exists a cross-over point between the full and infrastructure sharing cases. Full sharing provides better coverage at low \ac{SIR} values, as users experiencing strong interference are very susceptible to having unfavourable fading conditions, and, due to frequency-selective fading, coverage to such users may be significantly improved by simply transmitting over a number of spectrum bands. While high \ac{SIR} users also benefit from the aggregation of multiple bands, it is still more beneficial for them to simply change their point of attachment.

\begin{figure}[tb!]
\centering
	\includegraphics[width=.31\textwidth]{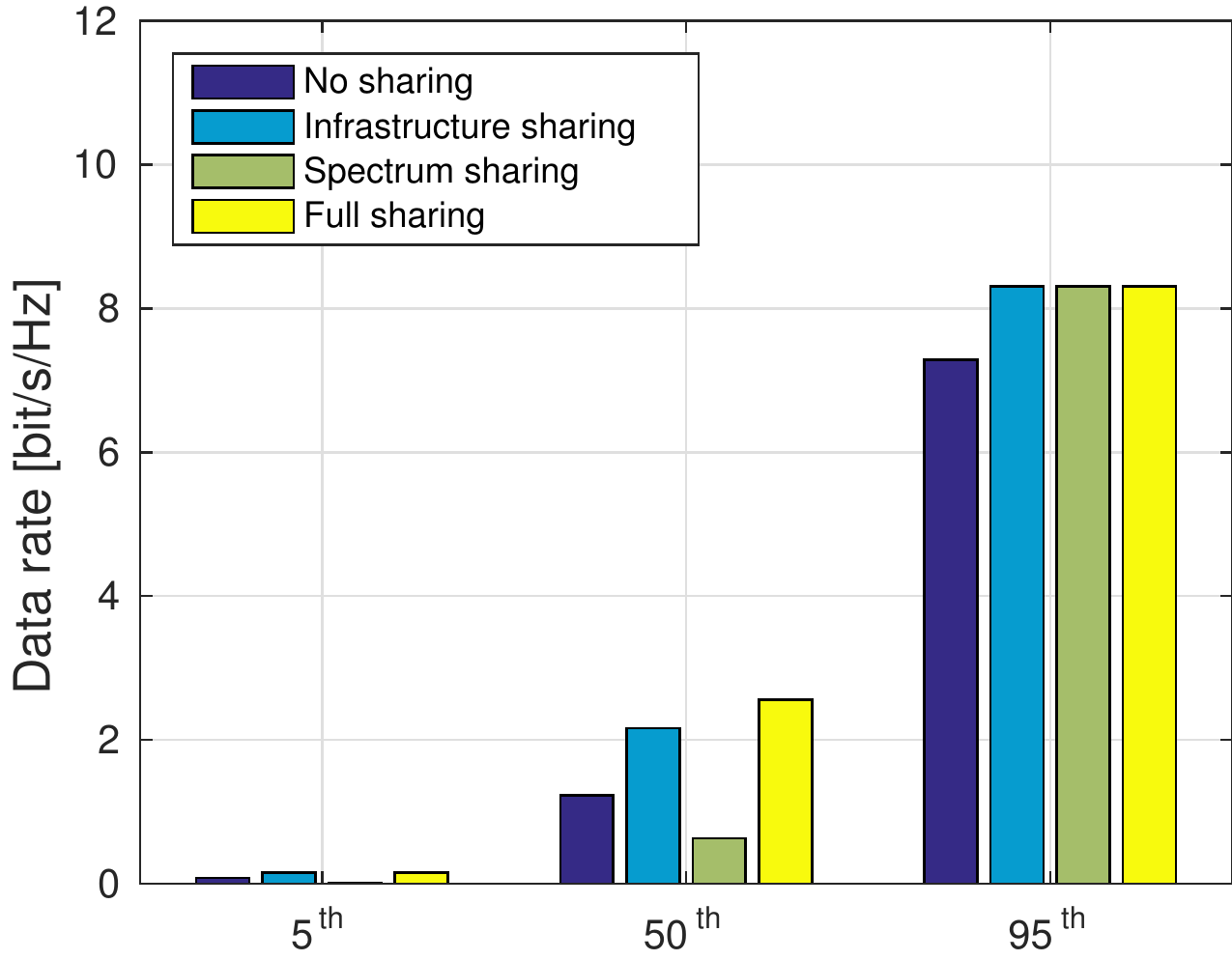}
		\vspace{-5mm}
	\caption{
	Data rate perceived by the $5^{\text{th}}$ percentile, median, and $95^{\text{th}}$ percentile users in different sharing scenarios when flat fading across shared spectrum bands is considered.
	}
	\label{fig:percentiles_ppp_corr}
\end{figure}

When the average user rate is considered, we see that full sharing (\unit[$4.30$]{bit/s/Hz}) significantly outperforms infrastructure (\unit[$3.11$]{bit/s/Hz}) and spectrum (\unit[$2.34$]{bit/s/Hz}) sharing. Surprisingly, unmanaged spectrum sharing among operators results in a rate that is only slightly better than that of a network where none of the resources are shared (\unit[$2.15$]{bit/s/Hz}). The reason for this is the increase in interference, which trumps the majority of the gains from pooling spectrum bands. When we consider user data rate at different percentiles (see \Fig{percentiles_ppp_corr}), we note that the rate distribution is highly skewed for each of the considered scenarios, as the median is significantly smaller than the average performance. This is especially true for the spectrum sharing scenario, where the average is $2.5$ times higher than the median. From \Fig{percentiles_ppp_corr}, we can also see that in each sharing scenario users that already enjoy good quality service ($95^{th}$) receive additional gains from sharing. These observations are consistent with observations in a recent paper \cite{BoccardiShokri-GhadikolaeiFodorEtAl_2016}, and signify the need for some form of coordination in shared spectrum use to more evenly distribute the sharing gains among users. While \Fig{percentiles_ppp_corr} shows only the case of flat fading across shared bands, based on our observations, similar conclusions can be drawn for the case of frequency-selective fading.

\subsection{Impact of the number of operators, and network density on coverage/rate}

Now, we look at the impact that the number of operators and the network density have on the performance of wireless network sharing.

\subsubsection{Number of sharing operators}

\begin{figure*}[tb!]
\centering
\subfigure[Infrastructure sharing \label{fig:Noperators_cp_inf}]{
 \includegraphics[width=0.31\textwidth]{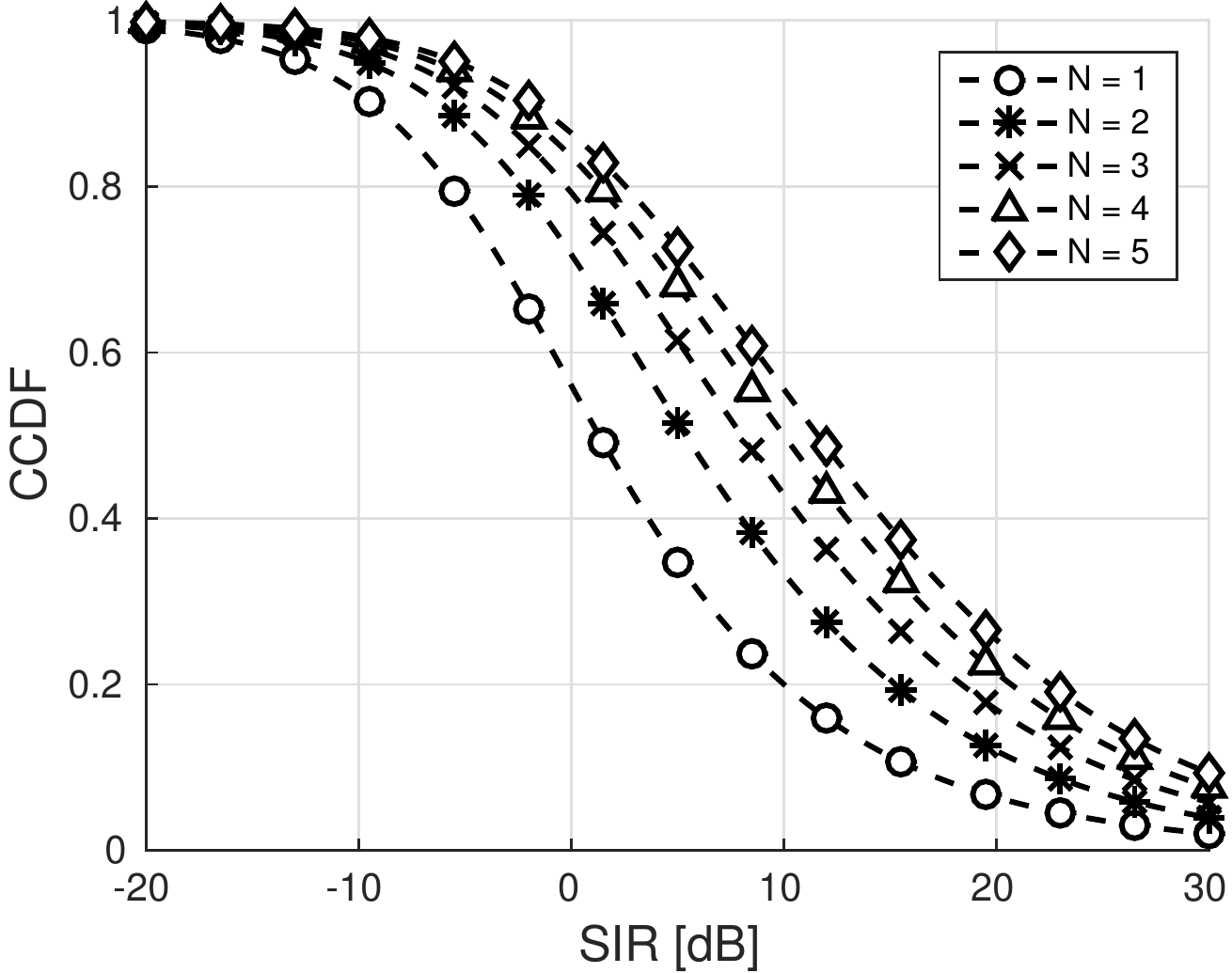}
}
\subfigure[Full sharing (frequency-selective fading) \label{fig:Noperators_cp_fs}]{
 \includegraphics[width=0.31\textwidth]{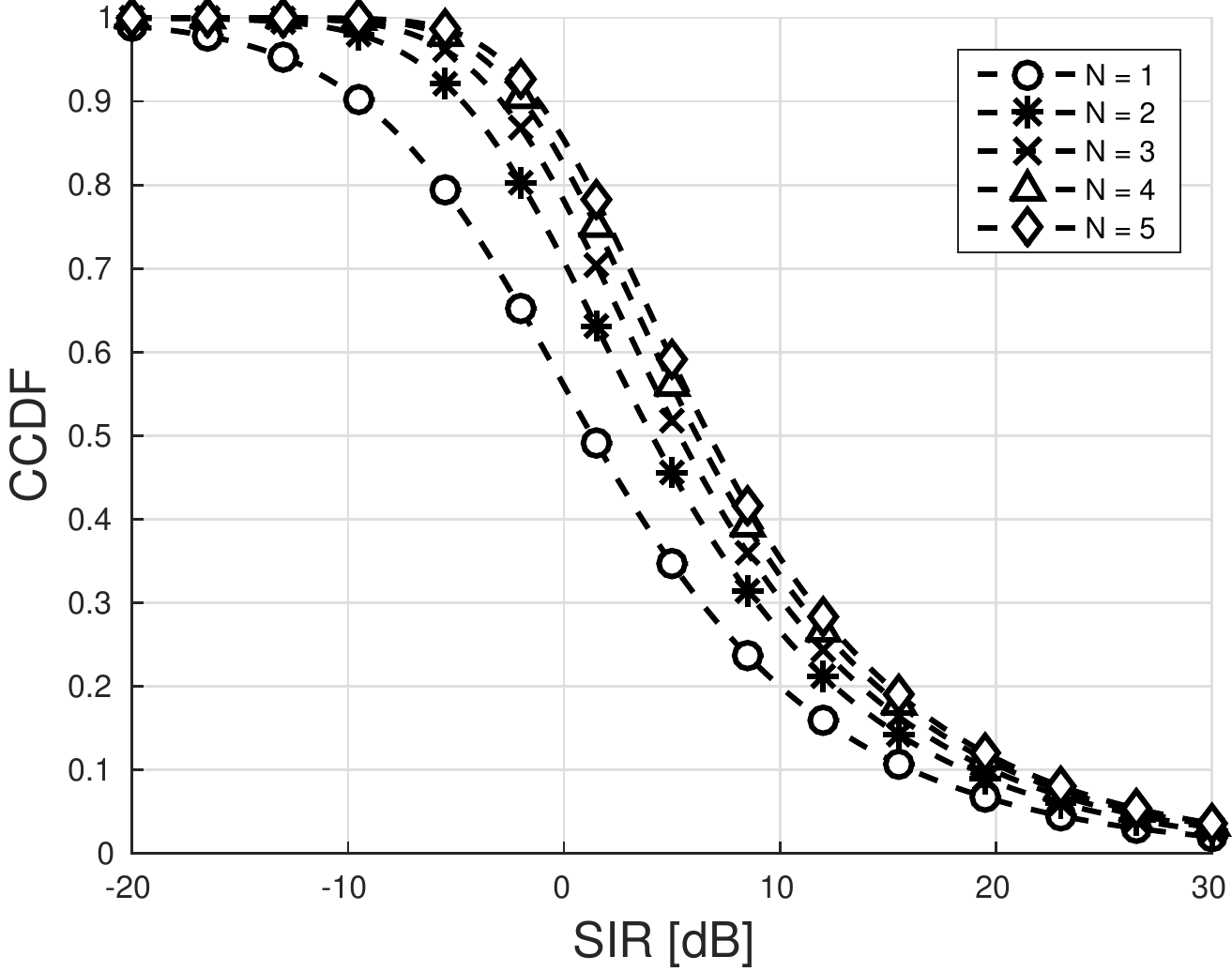}
 }
 
 \subfigure[Spectrum sharing (flat fading) \label{fig:Noperators_cp_ss_corr}]{
 \includegraphics[width=0.31\textwidth]{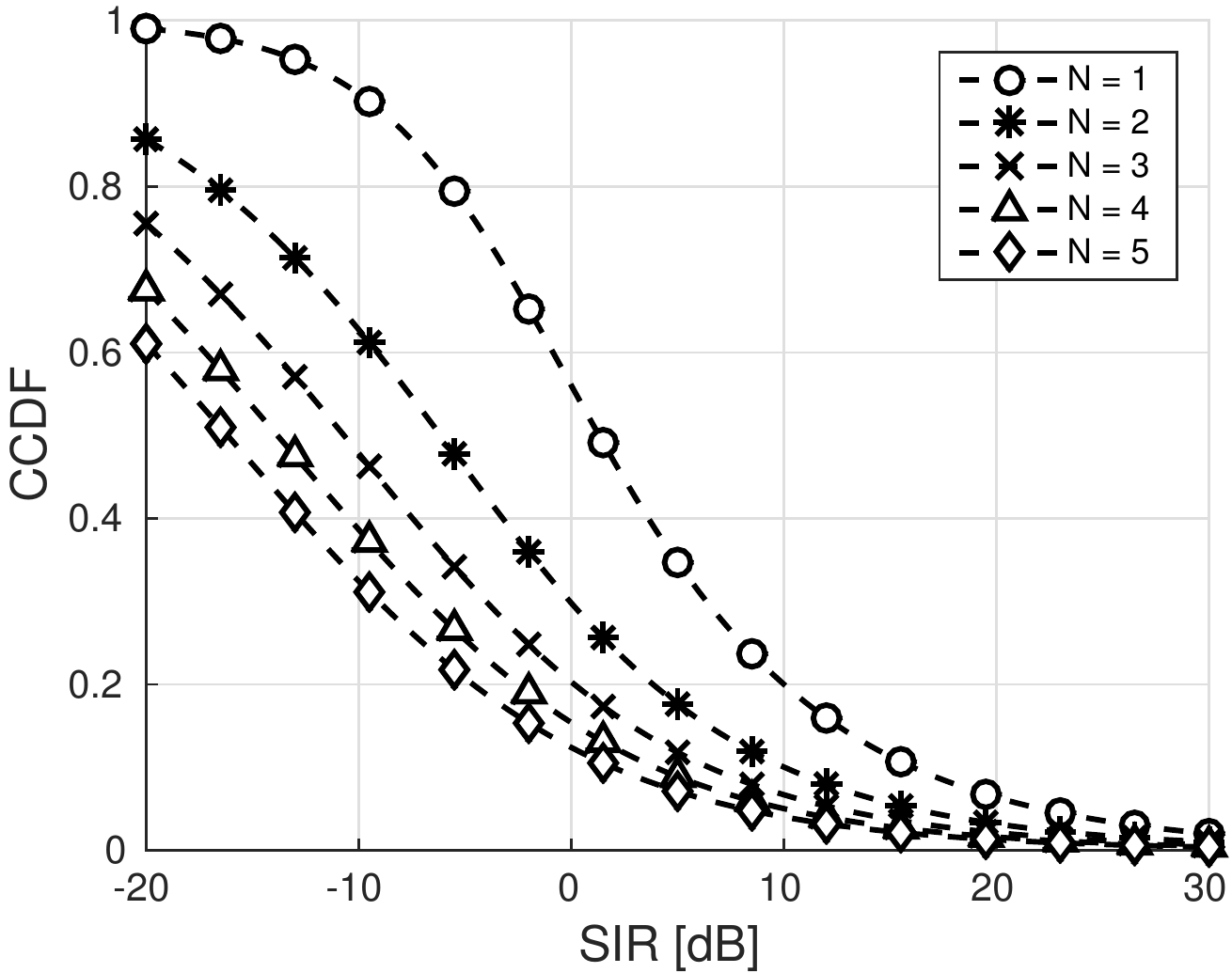}
}
\subfigure[Spectrum sharing (frequency-selective fading) \label{fig:Noperators_cp_ss_uncorr}]{
 \includegraphics[width=0.31\textwidth]{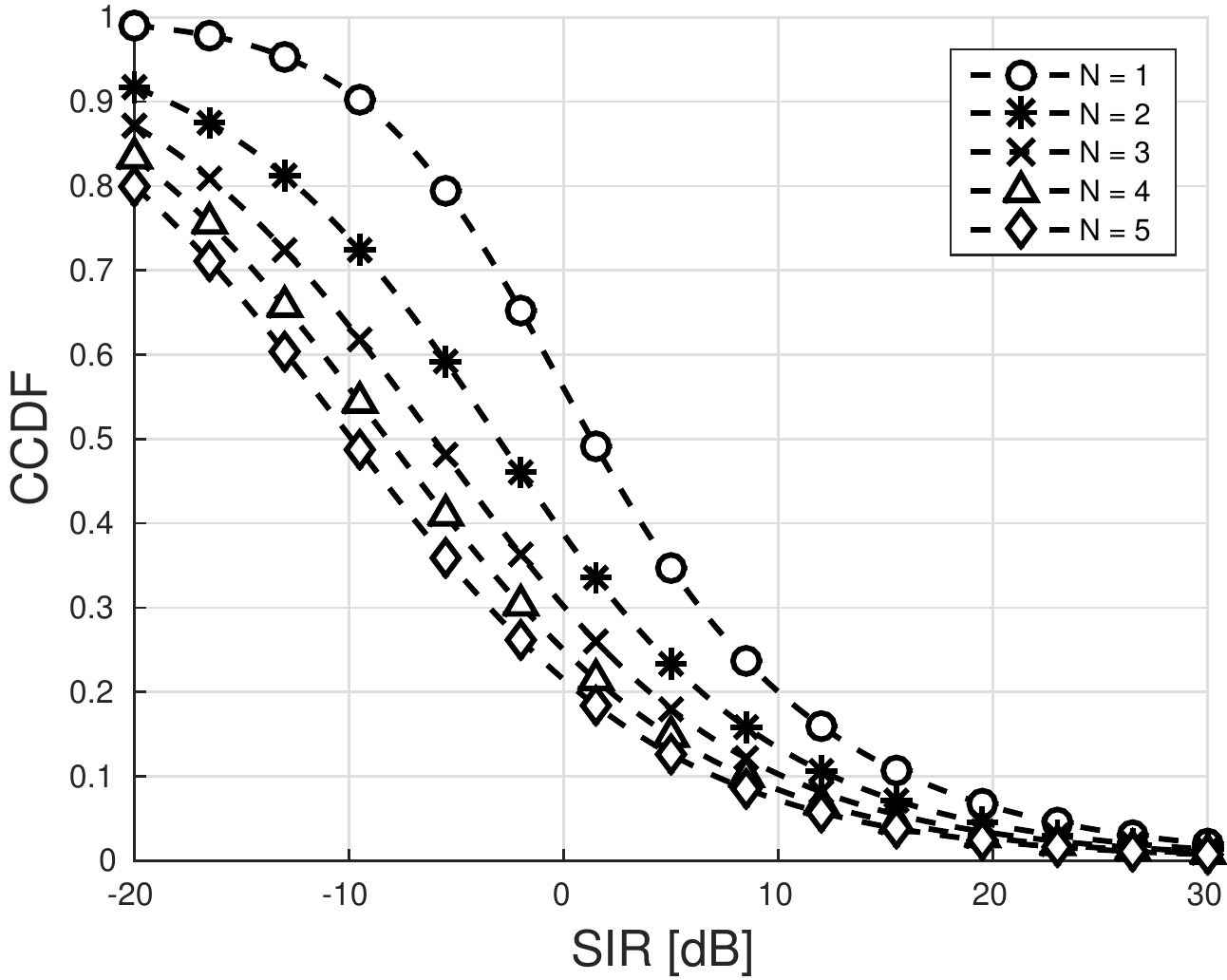}
 }
\caption{The impact of the number of sharing operators on the coverage probability.}
\label{fig:Noperators_cp}
\end{figure*}

\begin{figure}[tb!]
\centering
	\includegraphics[width=.31\textwidth]{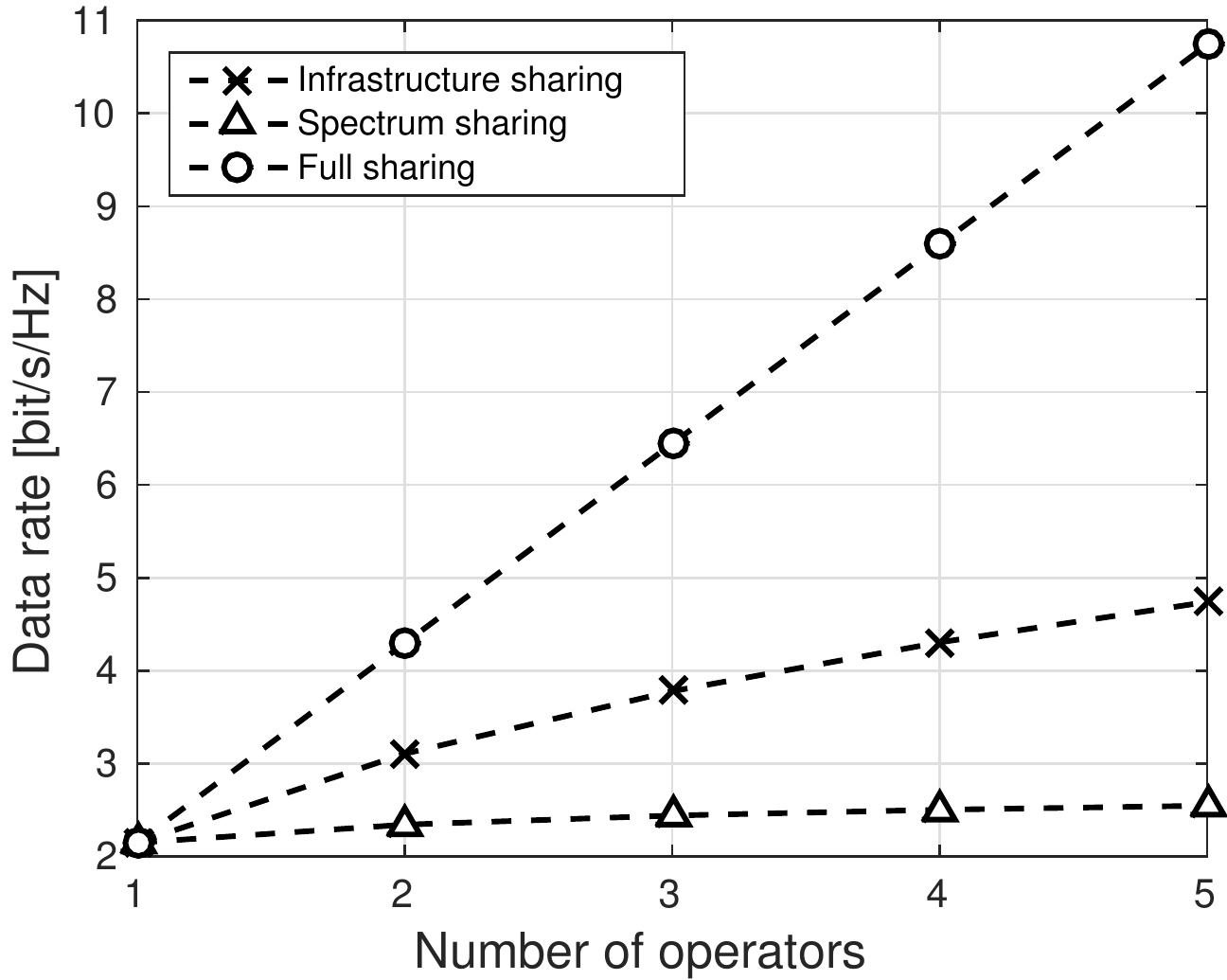}
		\vspace{-5mm}
	\caption{
	The impact of the number of sharing operators on the average user rate.
	}
	\label{fig:dr_Noperators}
\end{figure}

When infrastructure is shared (see \Fig{Noperators_cp_inf} and \Fig{Noperators_cp_fs}), coverage improves with the number of operators, with the greater gains coming from sharing with one additional operator, and diminishing marginal gains thereafter. However, when spectrum is shared without coordination, coverage decreases with the number of operators, due to an increase in the interference in the shared spectrum (see \Fig{Noperators_cp_ss_corr} and \Fig{Noperators_cp_ss_uncorr}). In each case, however, the average user rate improves (see \Fig{dr_Noperators}). While in the case of full sharing this rate grows linearly (as it is independent of the density\footnote{As shown in \cite{GaliottoEtAl_2014}, if distance-dependent shadowing is considered, there is a certain critical network density after which the spectral efficiency, and hence also the rate, is decreasing.}), sharing of only spectrum or infrastructure leads to much smaller gains.

\subsubsection{Network density imbalance}

\begin{figure*}[tb!]
\centering
\subfigure[Infrastructure sharing \label{fig:cp_inf_imb_majrev}]{
 \includegraphics[width=0.31\textwidth]{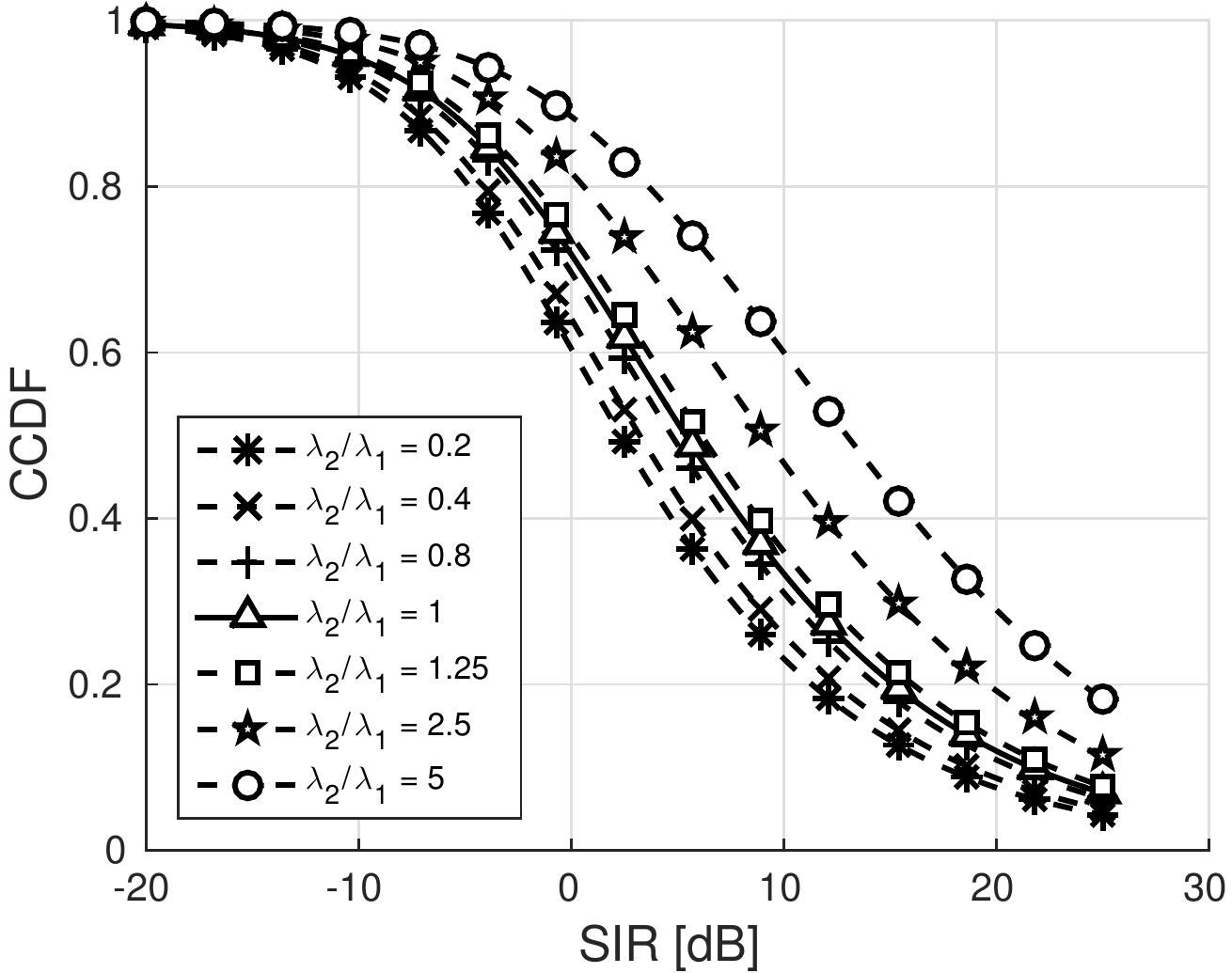}
}
\subfigure[Spectrum sharing 	(flat fading) \label{fig:cp_sp_imb_majrev}]{
 \includegraphics[width=0.31\textwidth]{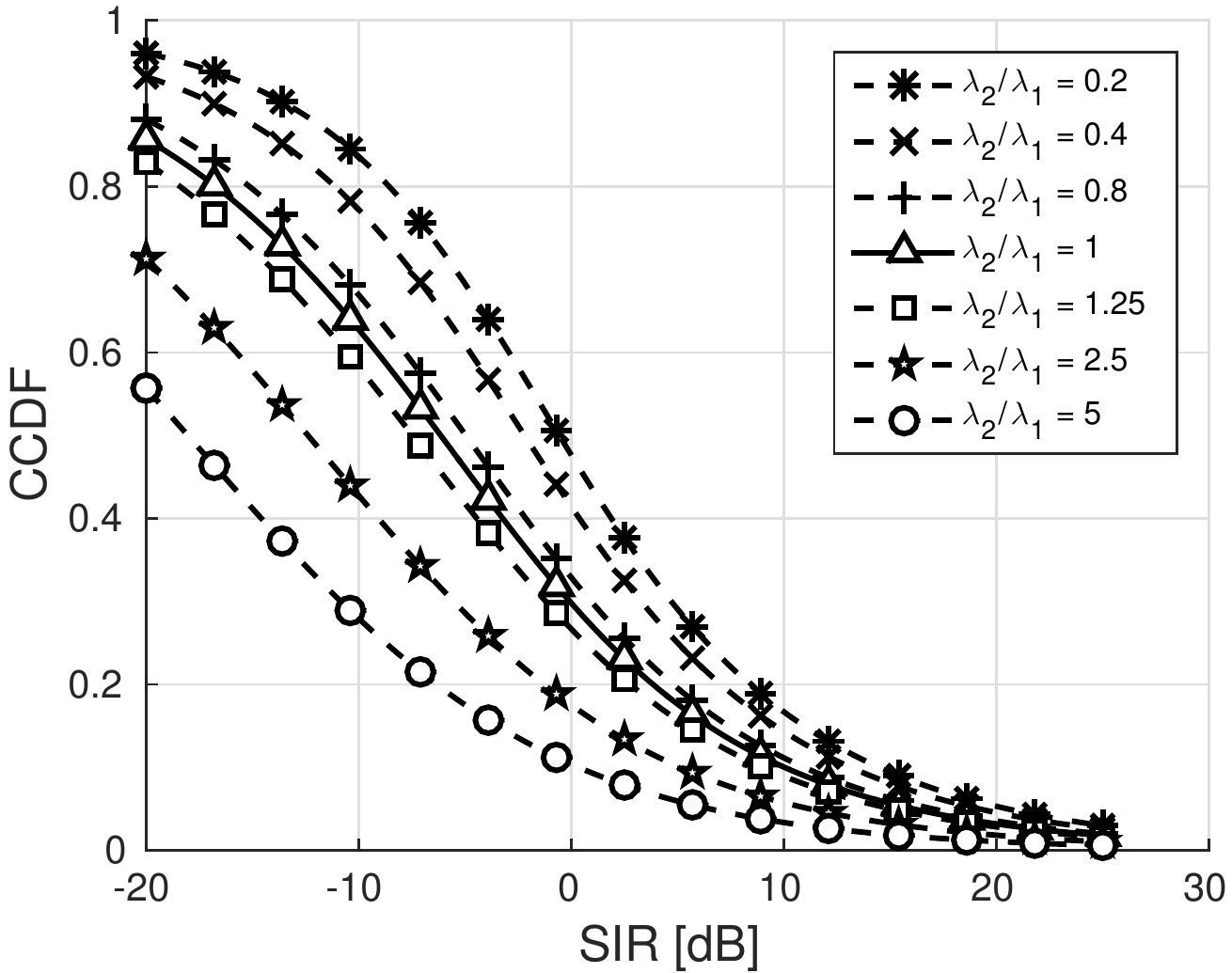}
 }
\caption{The impact of network density imbalance on the coverage probability, for a user of a network with density $\lambda_1$.}
\label{fig:inf_imb_majrev}
\end{figure*}

We define the network density imbalance as the ratio between the intensities of the shared networks $\lambda_2/\lambda_1$, with $\lambda_2/\lambda_1=1$ representing evenly sized networks. We can immediately note from \Fig{cp_inf_imb_majrev} that it is the user of a smaller operator ($\lambda_2/\lambda_1>1$) that benefits greatly, in terms of coverage probability, from infrastructure sharing. It comes from the fact that by adding on the infrastructure of a larger operator the distance to the nearest transmitter becomes significantly shortened. The opposite is true for spectrum sharing (see \Fig{cp_sp_imb_majrev}), whereby the user of the~larger operator experiences lower interference and therefore maintains better coverage than the user of the~smaller operator. These observations hold also if we extend our results to the case of spectrum sharing in bands experiencing frequency-selective fading.

\subsection{Impact of the small-scale fading}

Since we differentiate between sharing of spectrum bands with flat fading and frequency selective fading, it is important that we validate the impact of the fading model used.

\begin{figure*}[tb!]
\centering
\subfigure[Flat fading across shared bands\label{fig:cp_nakagami_correlated}]{
 \includegraphics[width=0.31\textwidth]{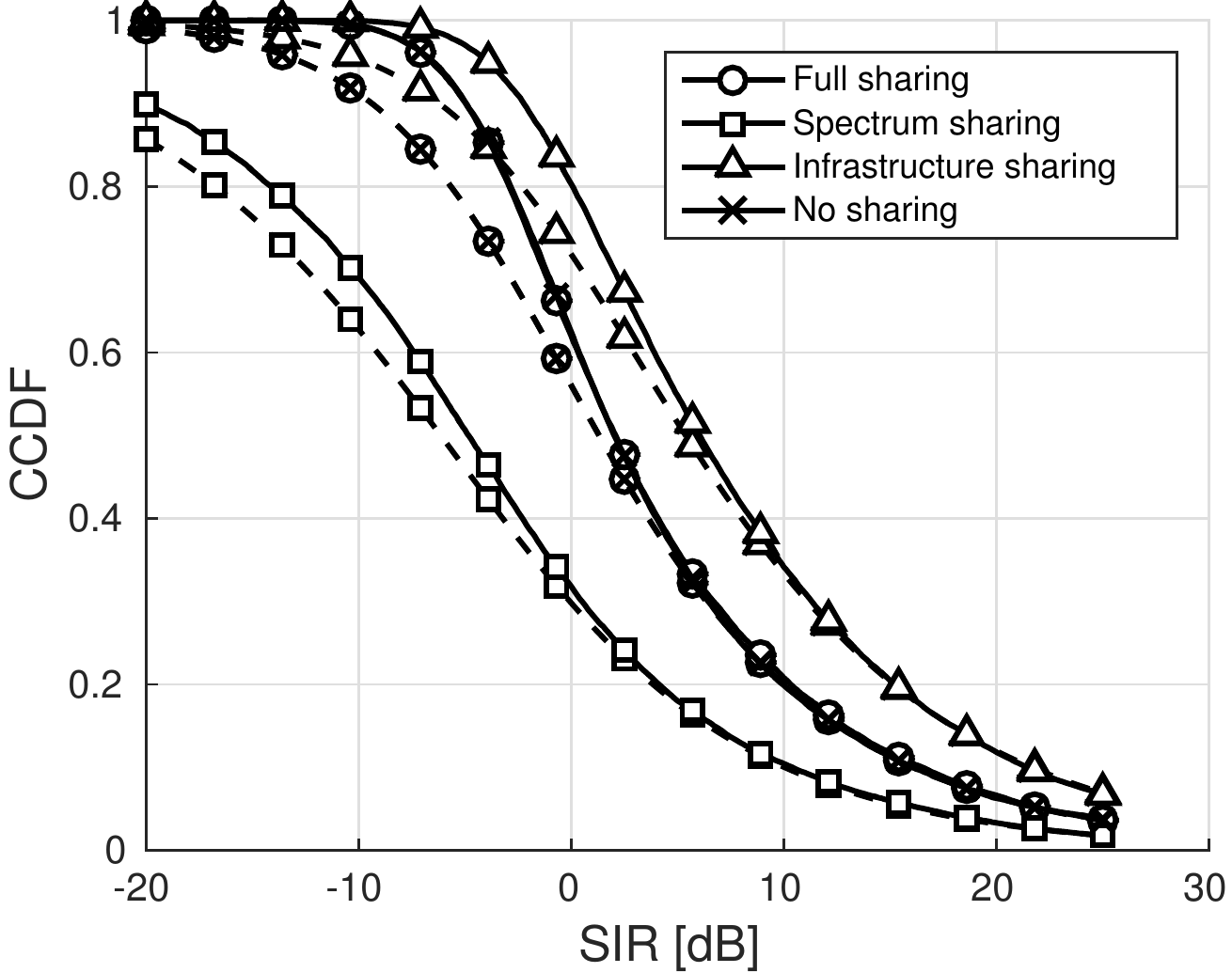}
}
\subfigure[Frequency-selective fading across shared bands\label{fig:cp_nakagami_uncorrelated}]{
 \includegraphics[width=0.31\textwidth]{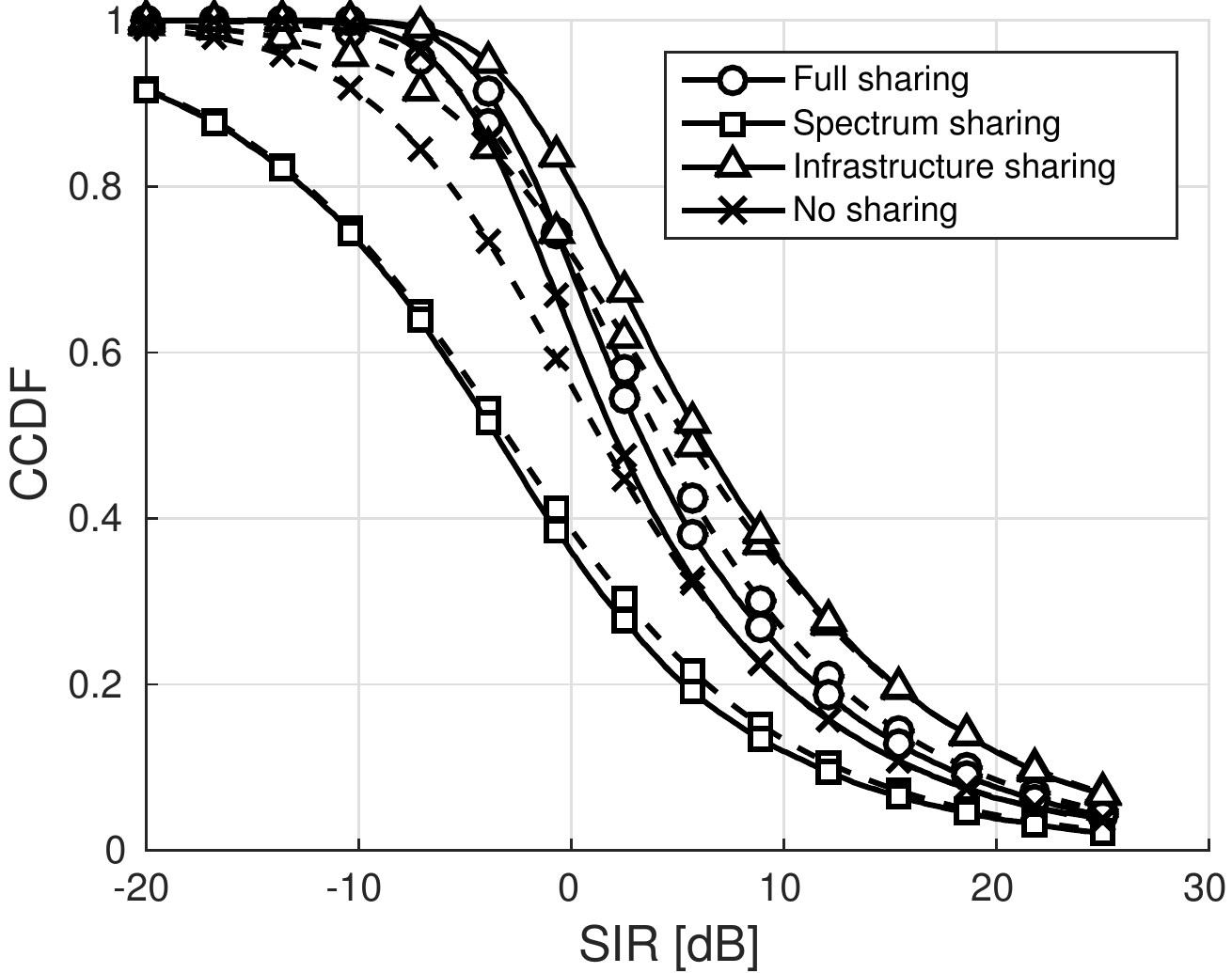}
 }
\caption{Comparison between coverage probability in various sharing scenarios with Rayleigh fading (dashed lines) and Nakagami-m fading (solid lines), with $m=5$.}
\label{fig:cp_nakagami}
\end{figure*}

As expected, Rayleigh fading, which is typically used to model channels with a strong multi-path component, yields a lower bound on the sharing performance when flat fading is assumed (see \Fig{cp_nakagami_correlated}). However, when we consider frequency selective fading (see \Fig{cp_nakagami_uncorrelated}), we see that Nakagami fading limits the gains obtained from spectrum sharing. This is because the variance of Nakagami fading decreases with the increase in the $m$ parameter, limiting gains available from having independent fading in the shared spectrum bands.

\subsection{Impact of spatial clustering}

\begin{figure*}[tb!]
\centering
\subfigure[Full sharing\label{fig:cp_clustering_full_correlated}]{
 \includegraphics[width=0.31\textwidth]{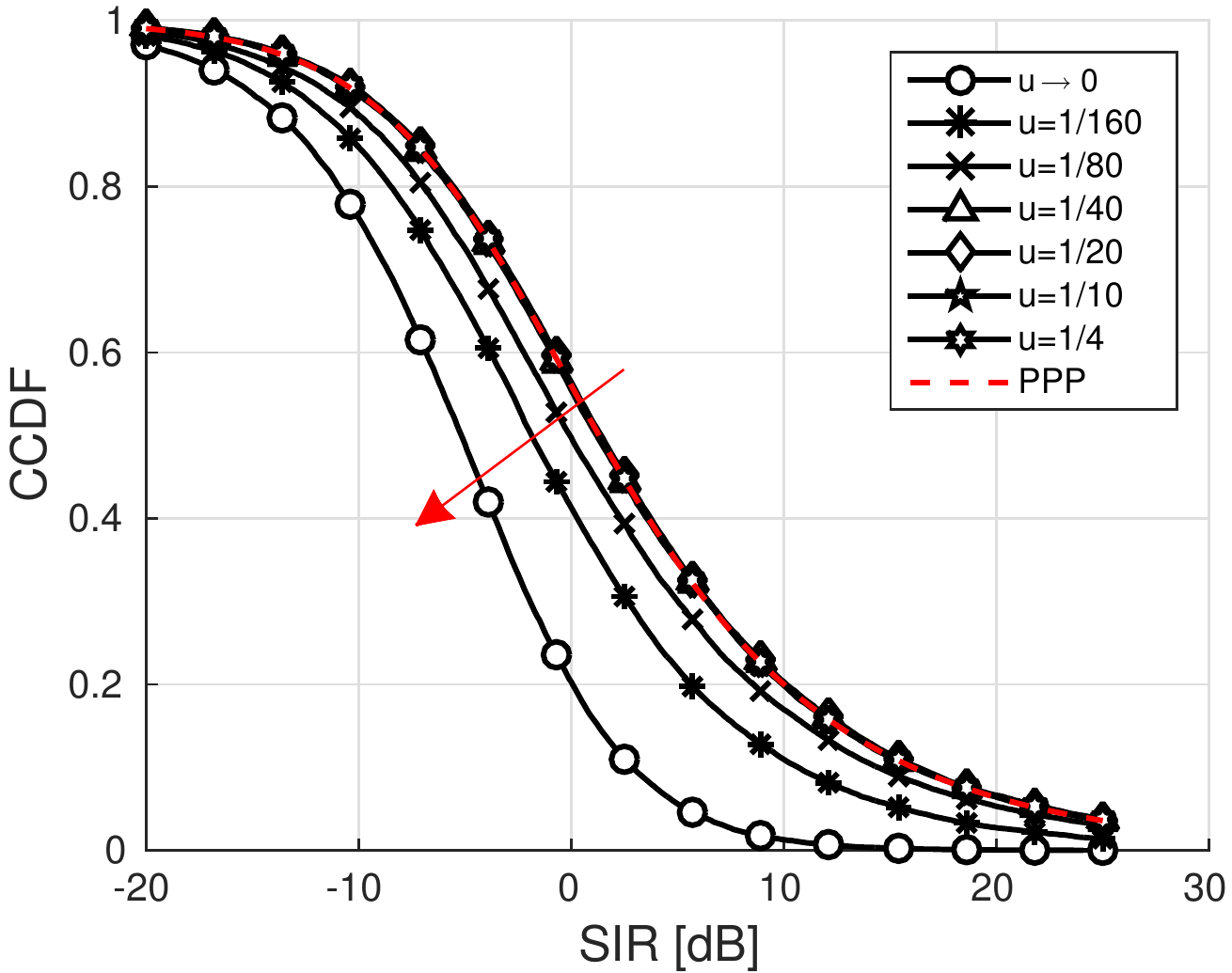}
}
\subfigure[Infrastructure sharing\label{fig:cp_clustering_roam_correlated}]{
 \includegraphics[width=0.31\textwidth]{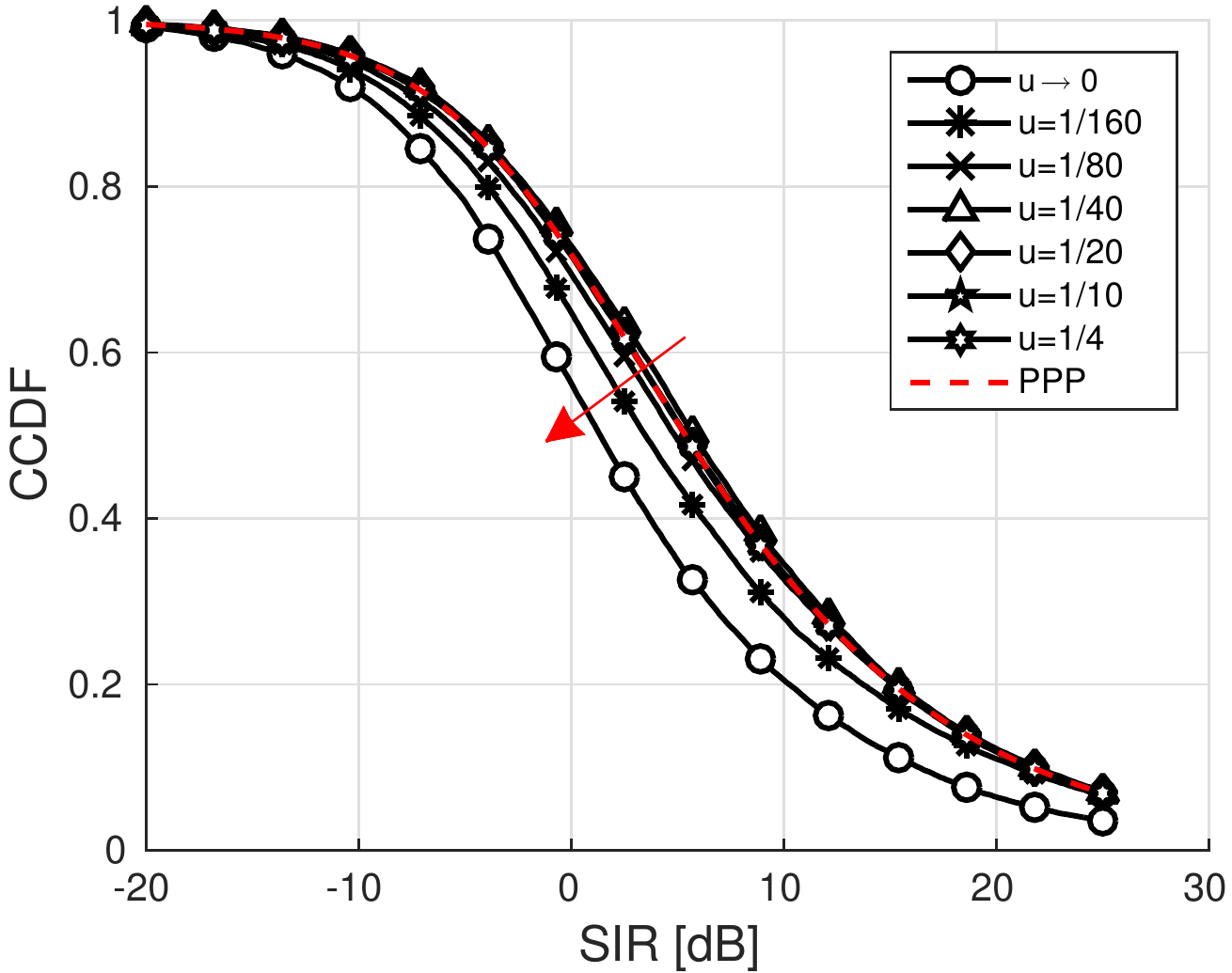}
}
\subfigure[Spectrum sharing\label{fig:cp_clustering_spec_correlated}]{
 \includegraphics[width=0.31\textwidth]{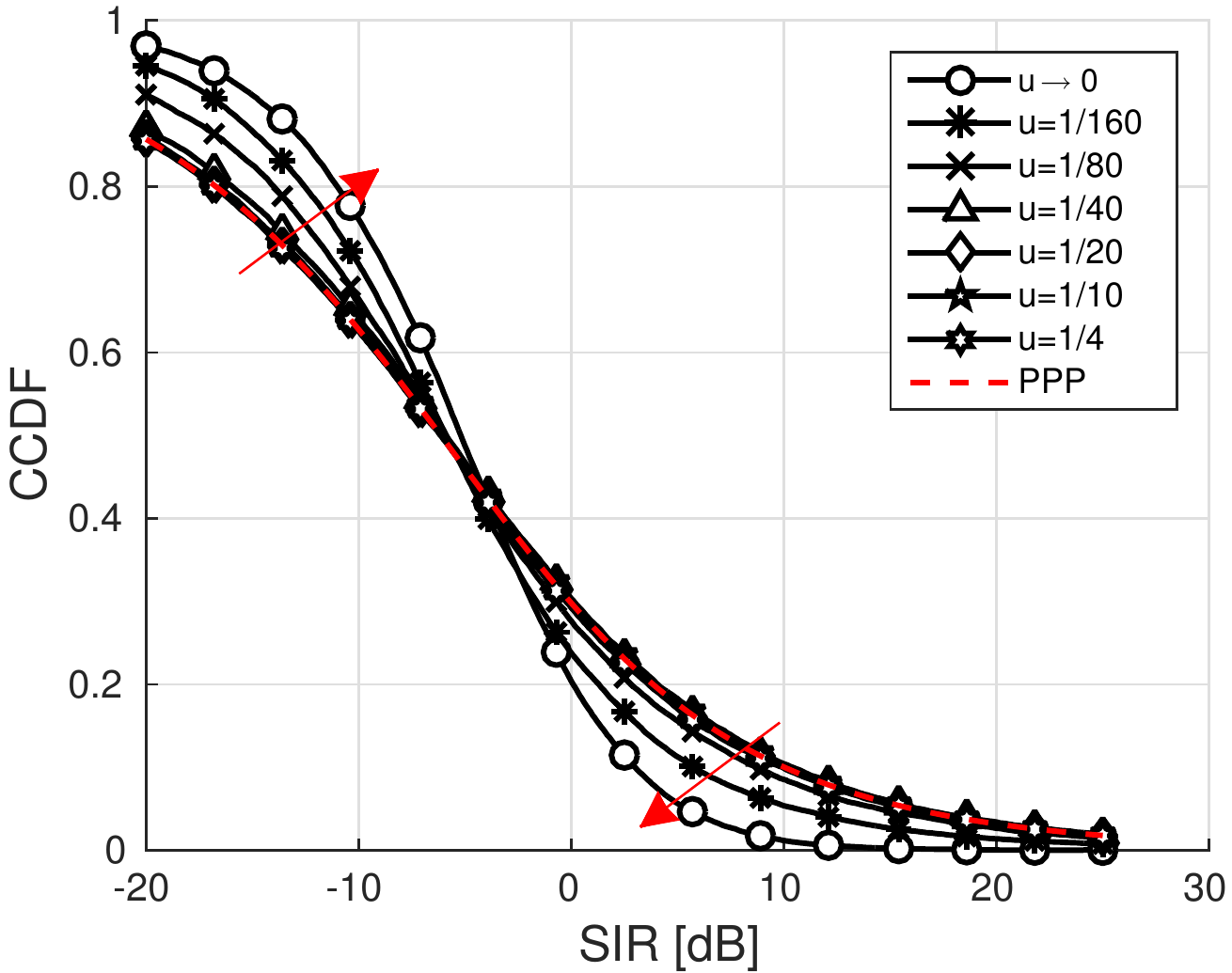}
}
\caption{
The coverage probability versus the relative cluster size in a shared clustered network (an independently distributed network is marked with a dashed line) when flat fading across shared bands is considered.
\vspace{-1mm}
}
\end{figure*}

We change our perspective and observe the impact that clustering of infrastructure has on different sharing scenarios. \Fig{cp_clustering_full_correlated}, \Fig{cp_clustering_roam_correlated}, and \Fig{cp_clustering_spec_correlated} present the coverage performance for each of our sharing scenarios when the infrastructure is: independently distributed (PPP), co-located ($u\rightarrow 0$), and clustered with a variable cluster radius normalized to the size of the simulation window (various values of $u$). 

As expected, in \Fig{cp_clustering_full_correlated} and \Fig{cp_clustering_roam_correlated} we observe that the increase in clustering (decreasing the cluster radius) deteriorates coverage (the direction of change marked with a red arrow) for both full and infrastructure sharing scenarios. For larger cluster radii the impact of clustering is negligible, while the closer the deployments get to co-location the more severe coverage deterioration becomes. In the case of spectrum sharing (\Fig{cp_clustering_spec_correlated}), the impact of clustering on the sharing performance is more complex. At low \ac{SIR} values we observe that the co-located infrastructure provides superior coverage while the independently distributed infrastructure fares the worst. After the cross-over point (slightly below \unit[$0$]{dB}) the situation reverses. This is related to an interplay between intra- and inter-cluster interference. Clearly, decreasing the cluster radius increases intra-cluster interference. However, as the cluster radius decreases so does the inter-cluster interference, which depends on the change of the distribution of distances to the interferers outside of the serving cluster (pairs of interferers are getting closer to each other). Effectively, while we observe that clustering reduces the probability of obtaining \ac{SIR}-values above the cross-over point (due to increased inter-cluster interference), we also observe an increase in the probability of obtaining \ac{SIR}-values below the cross-over point (due to reduced intra-cluster interference), as compared to the independently distributed infrastructure.

\begin{figure}[t]
\centering
	\includegraphics[width=.31\textwidth]{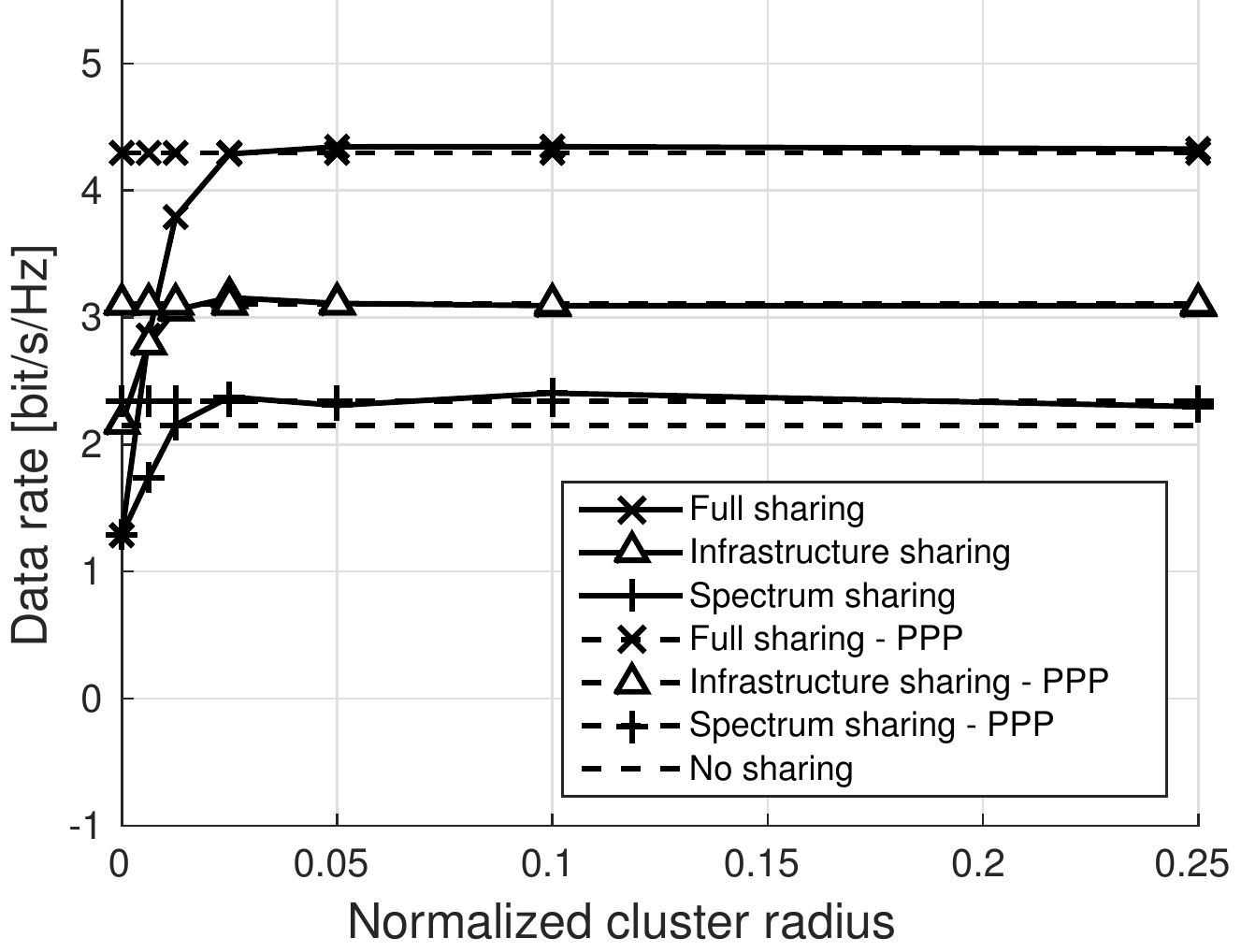}
	\caption{
	The average user rate versus the relative cluster size in a shared clustered network (an independently distributed network is marked with a dashed line).
	\vspace{-5mm}
	}
	\label{fig:rate_clustering_correlated}
\end{figure}

When average user rate is considered (\Fig{rate_clustering_correlated}), and the cluster radius is large enough, we can observe that the rate is always improved over the non-sharing scenario. Moreover, full sharing outperforms all the other sharing scenarios for most of the cluster radii. It is only at small cluster radii when the rate significantly drops and, when the infrastructure co-locates, full sharing achieves the same performance as spectrum sharing. From the figure we can also see that significant clustering and co-location of inter-operator networks have a dramatic effect on the rate performance of any scenario that relies on spectrum sharing (due to strong interference present in the shared network). These observations hold also if we extend our results to the case of spectrum sharing across bands experiencing frequency-selective fading.

\subsection{Impact of coordination in use of shared spectrum}

Finally, we perform a numerical study of the impact that a coordinated shared spectrum use has on inter-operator interference, and, effectively, the coverage and rate performance. In this subsection we rely on results obtained by means of \ac{MC} simulations, as the point process resulting from application of the coordinated spectrum use is a \ac{PHP} \cite{Haenggi_2013} which is considered to be analytically intractable\footnote{Typical approximations, which involve thinning a \ac{PPP} with a function that corresponds to the radius of the exclusion zone (see \cite{YazdanshenasanDhillonAfshangChong_2016}), do not provide a satisfactory level of accuracy.}. In addition to the normalized exclusion zone radius, we provide the percentage of transmitters that are not permitted to use the shared spectrum (they utilize only the spectrum band licensed to their operator). 

\begin{figure*}[tb!]
\centering
\subfigure[Spectrum sharing (flat fading)\label{fig:cpnum_ppp_5_0_correlated}]{
 \includegraphics[width=0.31\textwidth]{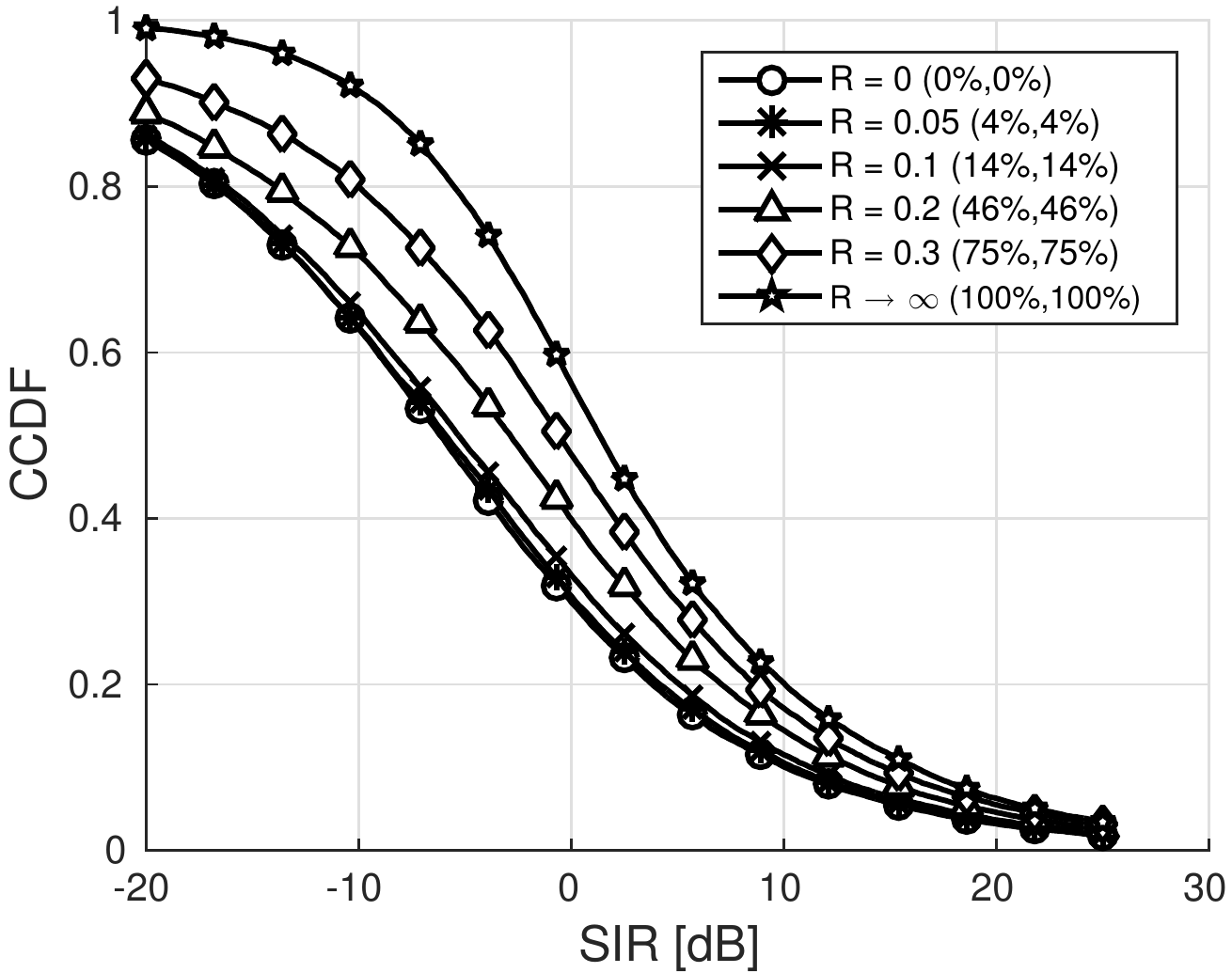}
}
\subfigure[Full sharing (frequency-selective fading)\label{fig:cpnum_ppp_0_0_correlated}]{
 \includegraphics[width=0.31\textwidth]{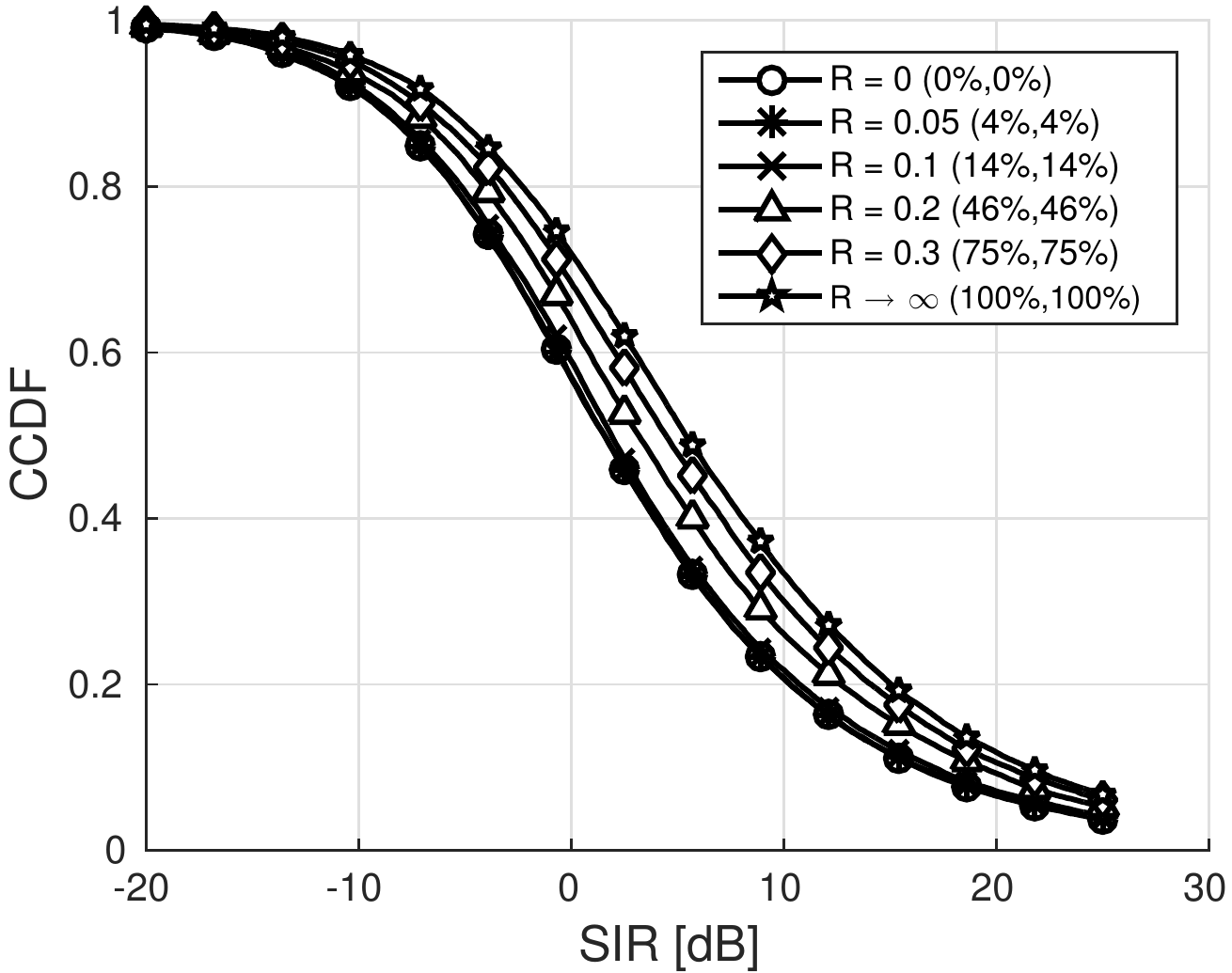}
}

\caption{
Coverage probability for scenarios with coordinated use of shared spectrum, where $R$ denotes the normalized exclusion zone radius and the percentages in brackets represent the portion of transmitters from each operator not allowed to utilize the spectrum of another operator.
}
\label{fig:cp-comparative-num}
\end{figure*}

When exclusion zones are applied evenly to each of the operators, we observe an improvement in the coverage performance (see \Fig{cpnum_ppp_5_0_correlated} and \Fig{cpnum_ppp_0_0_correlated}). The elimination of the nearest interferers ($R=0.1$) yields marginal improvements to coverage, and it is only if at least $50\%$ of the transmitters of the other operator are removed ($R>0.2$) that a more substantial improvement can be observed. Interestingly, the coverage gains for the full sharing scenario are much lower than in the spectrum sharing scenario. That is because in the full sharing scenario a user is always connected to the closest transmitter, therefore the interference it experiences is not as severe as in the spectrum sharing case. 

\begin{figure*}[tb!]
\centering
\subfigure[Spectrum sharing\label{fig:drnum_ppp_5_0_correlated}]{
 \includegraphics[width=0.3325\textwidth]{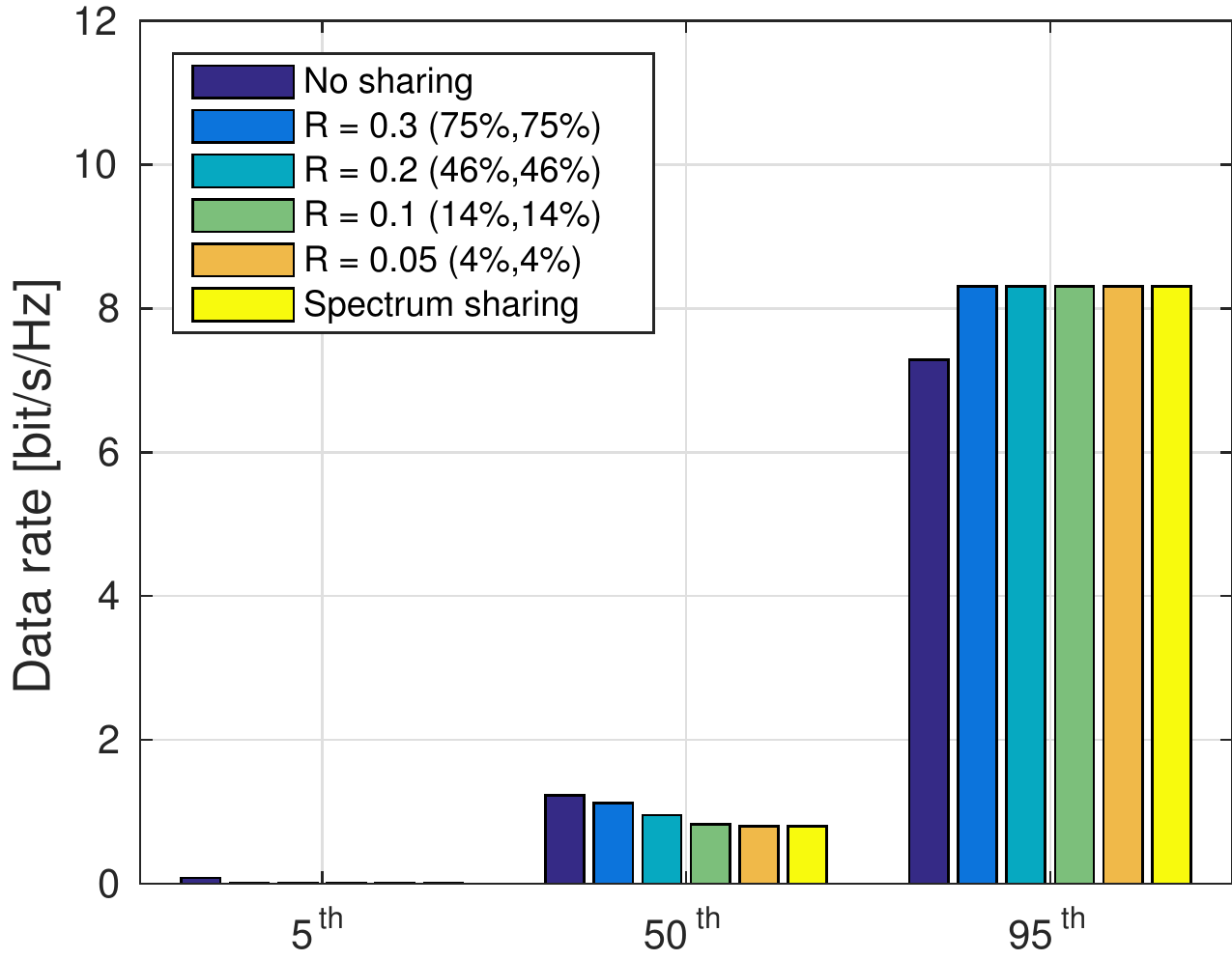}
}
~
\subfigure[Full sharing\label{fig:drnum_ppp_0_0_correlated}]{
 \includegraphics[width=0.31\textwidth]{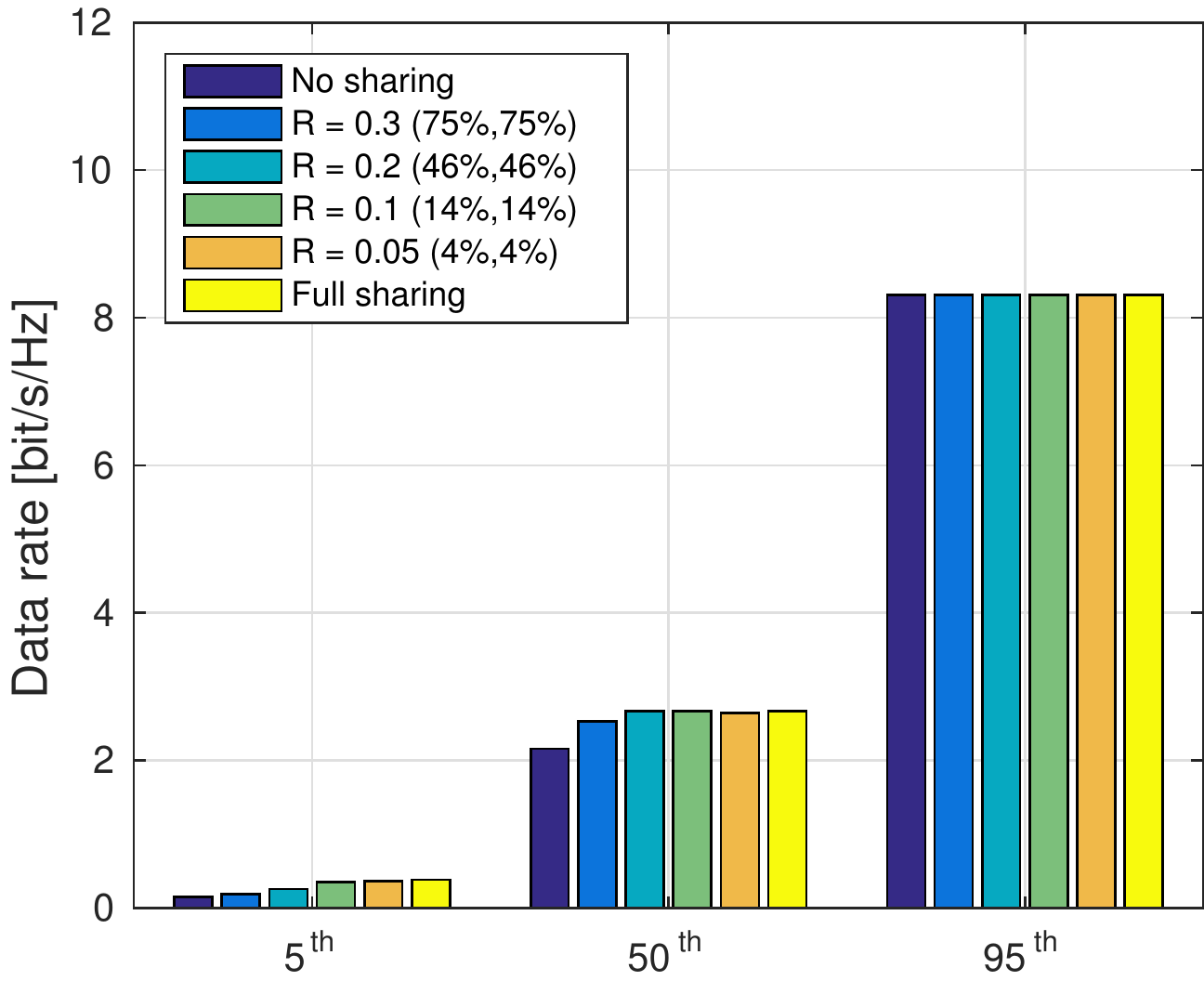}
}

\caption{
The rate performance of the $5^{\text{th}}$ percentile, median, and $95^{\text{th}}$ percentile users for scenarios with coordinated use of shared spectrum, where $R$ denotes the normalized exclusion zone radius and the percentages in brackets represent the portion of transmitters from each operator not allowed to utilize the spectrum of another operator.
}
\label{fig:dr_clustering-num}
\end{figure*}

Remarkably the improvement to coverage does not result in significant deterioration to the user rate. Imposing an exclusion zone that eliminates as much as $75\%$ of the transmitters has little effect on the average user rate. In order to explain this we need to understand that exclusion zones reduce interference (by removing interferers from the vicinity of the tagged transmitter), while reducing the chances that our tagged transmitter will enable a user to enjoy the benefits of spectrum sharing. These two effects cancel each other out, and we receive almost constant rate performance for a range of exclusion zone radii.
When we consider user rates at different percentiles (see \Fig{drnum_ppp_5_0_correlated} and \Fig{drnum_ppp_0_0_correlated}), we can see that while coverage improves the user rate does not deteriorate, with the median and $5^{\text{th}}$ percentile rates actually improving in the spectrum sharing scenario. Similar observations hold also for the case of frequency-selective fading, and spatial clustering in the shared infrastructure.

We applied the exclusion zone-based method of coordination simply because it allowed us to modify the geometry of the network, without changing our assumptions about the underlying transmission technology used. However, what we have seen is that real gains of spectrum sharing may only be available if more sophisticated methods of inter-operator coordination in shared spectrum use are applied. This is confirmed in \cite{BoccardiShokri-GhadikolaeiFodorEtAl_2016}, where application of inter-operator beamforming, combined with highly attenuating propagation of \ac{mmWave} bands, allows to bring up the rate gains across all users.

\section{Conclusion}
\label{sec:conclusion}

What we have shown in this work is that the performance gains brought about by mobile network resource sharing involve complex trade-offs that depend on what resources are being shared and on the nature of the mobile network deployments. The effects on the performance observed by a reference user include, among others, trade-offs in spectrum and radio access infrastructure sharing, and non-linear sharing gains with respect to the spatial clustering of the sharing networks or their sizes. In practical terms, this analysis allows us to capture how much sharing potential a network with specific spatial characteristics has or why a given sharing scenario may not be the most beneficial for some particular class of networks.

The goal of this work was to explore purely technical aspects of mobile network resource sharing in cellular bands, yet sharing decisions require also consideration of various non-technical aspects and externalities, such as operators' position in the market, their corporate strategy, legal framework surrounding mobile service provisioning, the political environment, or the economies of scope and scale for software-defined network hardware that may simplify managing of shared network resources. Some of these aspects were analyzed in other works (see, for example, \cite{Markendahl_2011,AhmedYangSungMarkendahl_2015}). However, more work needs to be done in order to develop a clear understanding of how these aspects can be quantified and accounted for to complement technical analysis such as ours.

Our work is well-grounded in the fundamentals of mobile network resource sharing in cellular bands, which are, for most technologies used today, interference-limited. Resource sharing in new spectrum bands, such as \ac{mmWave}, which offer better spatial separation between active transmitters, opens a possibility for definition of new resource sharing scenarios, and is already a subject of ongoing work efforts (see \cite{GuptaAndrewsHeath_2015,RebatoMezzavillaRanganZorzi_2016,GuptaAlkhateebAndrewsHeath_2016}).

%These capabilities go beyond the intuitive notion of distributing among the sharing parties the unused capacity and describe, among others, trade-offs in spectrum and radio access infrastructure sharing, and non-linear behaviours of the sharing gains with respect to the spatial clustering of the sharing networks or their sizes.

\section*{Acknowledgment}

This material is based in part upon work supported by the Science Foundation Ireland under grants no. 10/IN.1/3007 and 14/US/I3110.

\appendices

\section{Infrastructure sharing}
\label{app:infrastructure_sharing}

The superposition of \acp{PPP} is also a \ac{PPP}, with intensity equal to the sum of the intensities of the component processes. Effectively, in the infrastructure sharing scenario the point process that describes the serving transmitter is $\Phi$ with intensity $\lambda=\sum_{i\in\mathcal{N}}\lambda_i$. The \ac{pdf} of the distance to the closest transmitter in the Euclidean plane (obtained from \cite{Haenggi_2005}[Theorem 1]) can be expressed as:
\begin{equation}
f_{R}(r) = 2\pi \lambda r \exp(-\lambda\pi r^2).
\label{eq:pdf_r_ppp}
\end{equation} 

Now, in the infrastructure sharing scenario the interference to which the user is exposed depends on the network to which the user is connected (since each operator utilizes its own spectrum band). To find the coverage probability over all networks of all the operators we need to find the expectation over the coverage probability of each individual network, which requires calculating the probability of associating with the network of operator $n\in\mathcal{N}$. This association probability for a user with the network of operator $n$ in the infrastructure sharing scenario can be found as follows:
\begin{align}
\mathfrak{A}_{n} & = \mathbb{E}\Big[ \min_{j\in\mathcal{N}\setminus\{n\}} R_j > R_n \Big| R_n \Big] \nonumber \\
& = \mathbb{E}\Big[ \prod_{j\in\mathcal{N}\setminus\{n\}}\mathbb{P}\left( r < R_{j} \right) \Big| R_n = r \Big] \nonumber \\
& = \int^{\infty}_{0}\prod_{j\in\mathcal{N}\setminus\{n\}}\mathbb{P}\left( r < R_{j} \right) f_{R_{n}}(r)dr,
\label{eq:a_n_init}
\end{align}
where $f_{R_{n}}(r)$ is the distribution of distance from the reference user to the nearest transmitter of $n$, and $\mathbb{P}\left( r < R_{j} \right)$ may be interpreted as the probability that there is no transmitter of $j$ closer than a distance $r$ from the user; this can also be referred to as the null probability for network $j$, which is $\exp(-\pi\lambda_j r^2)$. After plugging the null probability for network $j$ and Eq.~(23) into Eq.~(24), we get that:
\begin{align}
\mathfrak{A}_{n} & = 2\pi \lambda_{n} \int^{\infty}_{0}r\exp\left(-\pi r^2 \sum_{j\in\mathcal{N}} \lambda_{j} \right)dr \nonumber \\
& = \frac{\lambda_n}{\sum_{j\in\mathcal{N}}\lambda_j}.
\label{eq:a_n_final}
\end{align}

Now, the coverage probability for the infrastructure sharing scenario can be expressed as (using the expression in Eq.~(2)):
\begin{equation}
p(\theta) = \sum_{i\in\mathcal{N}}\mathfrak{A}_{i} p(\theta | i), 
\end{equation}
where $p(\theta | i)$ is the coverage probability conditioned on connecting to the network of operator $i$, which can be expressed as:
\begin{equation}
p(\theta | i) = \int_{0}^\infty \exp(-\theta r^\alpha W) \mathcal{L}_{I_{i}}(\theta r^\alpha) f^\star_{R_i}(r)dr,
\end{equation}
where the Laplace transform $\mathcal{L}_{I_{i}}(\theta r^\alpha)$ of the interference in network $i$ is described using the form in Eq.~(3), and 
$f^\star_{R_i}(r)$ is the \ac{pdf} of the distance from a user to the nearest transmitter of network $i$, given that it is also its serving transmitter among all networks. The latter can be derived as follows:
\begin{align}
f^\star_{R_i}(r) & = \frac{dF^\star_{R_i}(r)}{dr} \nonumber \\
& = \frac{d}{dr} \bigg( 1 - \mathbb{P}\Big( R_{i} > r  \Big| \min_{j\in\mathcal{N}\setminus\{i\}} R_j > R_i \Big)\bigg) \nonumber \\
& \overset{(a)}{=}  - \frac{d}{dr}  \frac{\mathbb{P}\Big( R_{i} > r, \min_{j\in\mathcal{N}\setminus\{i\}} R_j > R_i \Big)}{\mathbb{P}\Big( \min_{j\in\mathcal{N}\setminus\{i\}} R_j > R_i \Big)} \nonumber \\
& \overset{(b)}{=} \frac{1}{\mathfrak{A}_i} \frac{d}{dr} \int^{r}_{-\infty}\prod_{j\in\mathcal{N}\setminus\{i\}}\mathbb{P}\left( R_{j} > r \right) f_{R_{i}}(r)dr \nonumber \\
& = \frac{ 2\pi \lambda_i}{\mathfrak{A}_i} r\exp\Big( -\pi r^2\sum_{j\in\mathcal{N}}\lambda_j \Big),
\end{align}
where $F^\star_{R_i}(r)$ denotes the cumulative distribution function; (a) the denominator was derived in Eq.~(25); (b) the inner probability is the null probability of network $j$, i.e., $\exp(-\pi\lambda_j r^2)$. Finally, after making the necessary substitutions, we get the following expression:
%\begin{align}
%p(\theta) & =  2\pi \sum_{i\in\mathcal{N}}\lambda_i \int_{0^+}^{\infty} \exp\Big(-\theta r^{\alpha} W\Big)\cdot\\\nonumber
%& \exp\bigg(-2\pi r^2 \Big(\sum_{j\in\mathcal{N}}\lambda_j + \lambda_i \mathfrak{Z}(\theta,\alpha) \Big)\bigg) rdr.
%\label{eq:pc_infrastructure_sharing_general}
%\end{align}
\begin{align}
p(\theta) & =  2\pi \sum_{i\in\mathcal{N}}\lambda_i \int_{0}^{\infty} \exp\Big(-\theta r^{\alpha} W\Big)\nonumber\\
& \exp\bigg(-2\pi r^2 \Big(\sum_{j\in\mathcal{N}}\lambda_j + \lambda_i \mathfrak{Z}(\theta,\alpha) \Big)\bigg) rdr.
\label{eq:pc_infrastructure_sharing_general}
\end{align}

The derivation of the average user rate requires plugging the obtained formula into Eq.~(5), which concludes the proof.

\section{Sharing spectrum bands under flat fading}
\label{app:spectrum_sharing}

Let us observe that the point process describing the closest transmitter to a reference user of operator $n$ is $\Phi_n$ with intensity $\lambda_n$. Following the same logic as previously, the expression for the probability of finding the closest transmitter at distance $r$ is the same as in Eq.~(23), yet with intensity $\lambda_n$.

In the spectrum sharing scenario the interference to a user of operator $n$ comes from: (i) all the transmitters of operator $n$ that are at a distance $r$ and further apart, which we denote as $I_r(\Phi_n)$, and (ii) all the transmitters of all operators other than $n$, which may be at any arbitrary small distance to the reference user, which we denote as $I_0(\Phi_j)$ and $j\neq n$. Now, the total interference is $I=I_r(\Phi_n)+\sum_{j\neq n}I_0(\Phi_j)$, and its Laplace transform can be derived as follows:
\begin{align}
\mathcal{L}_{I}(s) & = \mathbb{E}_{I}\left[\exp(-sI) \right] \nonumber \\
& = \mathbb{E}_{I}\Big[\exp(-sI_r(\Phi_n)) \exp(-s\sum_{j\neq n}I_0(\Phi_j))\Big] \nonumber \\
& \overset{(a)}{=} \mathbb{E}_{I_r(\Phi_n)}\Big[\exp(-sI_r(\Phi_n))\Big] \nonumber \\
& \prod_{j\neq n}\mathbb{E}_{I_0(\Phi_j)}\Big[\exp(-s\sum_{j\neq n}I_0(\Phi_j))\Big] \nonumber \\
& = \mathcal{L}_{I_r(\Phi_n)}(s)\prod_{j\neq n}\mathcal{L}_{I_0(\Phi_j)}(s),
\label{eq:laplace_ppp_spec}
\end{align}
where $s=\theta r^\alpha$, and (a) follows from the worst case scenario interference assumption, which allows us to treat the interference coming from the the networks of different operators as independent, and therefore replace the expectation of the product with the product of individual expectations. 

Now, $\mathcal{L}_{I_r(\Phi_n)}(s)$ is given in Eq.~(3), while $\mathcal{L}_{I_0(\Phi_j)}(s)$ can be obtained as follows:
\begin{align}
\mathcal{L}_{I_0(\Phi_j)}(\theta r^\alpha) & = \exp\Big(-\pi r^2 \lambda_j \theta^{2/\alpha}\int^{\infty}_{0}\frac{1}{1+u^{\alpha/2}}du \Big) \nonumber \\
& = \exp\Big(-\pi r^2 \lambda_j \mathfrak{Z}_0(\theta,\alpha) \Big),
\label{eq:laplace_ppp_spec_0}
\end{align}
where $\mathfrak{Z}_0(\theta,\alpha) = \theta^{2/\alpha}\Gamma(1+2/\alpha)\Gamma(1-2/\alpha)$. Finally, when we plug Eq.~(3) and Eq.~(31) into Eq.~(30), and, subsequently, to Eq.~(2), we get the expression for the coverage probability of operator $n$ as in Eq.~(16).
%\begin{equation}
%p_n(\theta) =  \pi\lambda_n \int_{0^+}^{\infty} \exp(-\theta v^{\alpha/2} W)\exp(-\lambda_n \pi v)\exp\bigg(-\pi v\Big(\lambda_n\mathfrak{Z}(\theta,\alpha) + \mathfrak{Z}_0(\theta,\alpha)\sum_{j\neq n}\lambda_j\Big)\bigg) dv.
%\label{eq:pc_spectrum_sharing_general}
%\end{equation}

Calculation of the average user rate across the shared spectrum requires observing that the expectation of a sum of random variables is a sum of expectations of each individual random variable. Hence, we get that the average user rate is as in Eq.~(14), where $p_n(\cdot)$ is the coverage probability from Eq.~(16), which concludes the proof.

\section{Sharing spectrum bands under frequency-selective fading}
\label{app:spectrum_sharing_diversity}

When spectrum bands experiencing frequency-selective fading are shared, the coverage probability of a reference user of operator $n$ in the spectrum sharing scenario can be obtained as follows:
\begin{align}
p_{n} (\theta) & \overset{(a)}{=} 1 - \mathbb{P} \bigg(\bigcap_{i\in\mathcal{N}} \mathrm{SINR}_i \leq \theta\bigg) \nonumber \\
& = 1 - \mathbb{P} \bigg(\bigcap_{i\in\mathcal{N}} h_x^i \leq \theta/l(x)\left( W + I_{i} \right)\bigg) \nonumber \\
& \overset{(b)}{=} 1 - \mathbb{E}_{I}\bigg[\prod_{i\in\mathcal{N}} \Big(1 - \exp\Big(-s \big(W + I_i\big)\Big)\Big) \bigg] \nonumber \\
& \overset{(c)}{=} 1 - \mathbb{E}_{I}\bigg[\bigg(1 - \exp\Big(-s W\Big)\nonumber\\
&\prod_{y\in\Phi_n\setminus\{x\}}\frac{1}{\Big(1+s l(y)\Big)}\prod_{j\in\mathcal{N}\setminus\{n\}} \prod_{y\in\Phi_j}\frac{1}{\Big(1+s l(y)\Big)}\bigg)^{|\mathcal{N}|}\bigg] \nonumber \\
& \overset{(d)}{=} \sum_{k\in\mathcal{N}} \binom{|\mathcal{N}|}{k} (-1)^{k+1} \exp\Big(-s kW\Big)\mathbb{E}_I\bigg[\nonumber\\
&\prod_{y\in\Phi_n\setminus\{x\}}\frac{1}{\Big(1+s l(y)\Big)^k} \prod_{j\in\mathcal{N}\setminus\{n\}} \prod_{y\in\Phi_j}\frac{1}{\Big(1+s l(y)\Big)^k}\bigg] \nonumber \\
& \overset{(e)}{=} \sum_{k\in\mathcal{N}} \binom{|\mathcal{N}|}{k} (-1)^{k+1} \exp\Big(-s kW\Big)\mathcal{L}^k(s,\lambda_n)\nonumber\\
&\prod_{j\in\mathcal{N}\setminus\{n\}}\mathcal{L}^k_0(s,\lambda_j),
\label{eq:pc_n_proof_p1}
\end{align}
%& = 1 - \mathbb{E}_I\bigg[\prod_{i\in\mathcal{N}} \Big(1 - \exp\Big(-s W\Big) \prod_{y\in\Phi_n\setminus\{x\}}\Big(\frac{1}{1+s l(y)}\Big) \prod_{j\in\mathcal{N}\setminus\{n\}} \prod_{y\in\Phi_j}\Big(\frac{1}{1+s l(y)}\Big)\bigg] \nonumber \\

where (a) conditioned on the distance $r$ to the nearest transmitter in $x$, the probability of coverage from at least one spectrum band may be expressed in terms of the probability of outage in each shared spectrum band; (b) $s = \theta/l(x)$; (c) given Rayleigh fading; (d) applying binomial expansion; (e) using the expression for the probability generating functional of the PPP \cite{Haenggi_2013}[Theorem 4.9]. Now, $\mathcal{L}^k(s,\lambda_n)$ can be found as follows:
\begin{align}
&\mathcal{L}^k(s,\lambda_n) \overset{(a)}{=} \exp\bigg(-2\pi\lambda_n\int_r^{\infty} \Big( 1 - \Big(\frac{1}{1 + s v^{-\alpha}}\Big)^k\Big)v dv \bigg) \nonumber \\
& = \exp\bigg(-2\pi\lambda_n\int_r^{\infty} \Big( 1 - \Big(1 - \frac{1}{1 + v^{\alpha}/s}\Big)^k\Big)v dv \bigg) \nonumber \\
& \overset{(b)}{=} \exp\bigg(-\frac{2\pi\lambda_n}{\alpha}\int_{r^\alpha}^{\infty} \Big( 1 - \Big(1 - \frac{1}{1 + u/s}\Big)^k\Big)u^{2/\alpha-1}du \bigg) \nonumber \\
& \overset{(c)}{=} \exp\bigg(-\frac{2\pi\lambda_n}{\alpha} \sum^{k}_{l=1} \binom{k}{l} (-1)^{l+1} \int_{r^\alpha}^{\infty} \frac{u^{2/\alpha-1}}{\big(1 + u/s\big)^l}du \bigg) \nonumber \\
& \overset{(d)}{=} \exp\bigg(-\pi\lambda_n  r^2 \sum^{k}_{l=1} \binom{k}{l} (-1)^{l+1} \mathfrak{Z}(\theta,\alpha,l) \bigg),
\label{eq:laplace_k_n_spec}
\end{align}
%& = \exp\bigg(-2\pi\lambda_n\int_r^{\infty} \Big( 1 - \Big(\frac{1}{1 + s v^{-\alpha}} + 1 - 1\Big)^k\Big)v dv \bigg), \nonumber \\
%& = \exp\bigg(-2\pi\lambda_n\int_r^{\infty} \Big( 1 - \Big(1 - \frac{s v^{-\alpha}}{1 + s v^{-\alpha}}\Big)^k\Big)v dv \bigg), \nonumber \\
%& = \exp\bigg(-2\pi\lambda_n\int_{r^\alpha}^{\infty} \Big( 1 - \Big(1 - \frac{1}{1 + u/s}\Big)^k\Big)\frac{u^{2/\alpha-1}}{\alpha}du \bigg), \nonumber \\
where (a) $v=l(y)$ and switching to polar coordinates; (b) $u=v^\alpha$; (c) applying binomial expansion; (d) $\mathfrak{Z}(\theta,\alpha,l) = \frac{2\theta^{l}}{l\alpha - 2}\mbox{$_2$F$_1$}(l, l-2/\alpha;\,l-2/\alpha+1;\,-\theta)$, based on \cite{GradshteynRyzhik_2007}[Equation 3.194.2]. Following the same logic as above, we find $\mathcal{L}^k_0(s,\lambda_j)$ as:
\begin{equation}
\mathcal{L}^k_0(s,\lambda_j) = \exp\bigg(-\pi\lambda_j  r^2 \sum^{k}_{l=1} \binom{k}{l} (-1)^{l+1} \mathfrak{Z}_0(\theta,\alpha,l) \bigg),
\label{eq:laplace_k_0_n_spec}
\end{equation}
where $\mathfrak{Z}_0(\theta,\alpha,l) =  \frac{2\theta^{2/\alpha}}{\alpha} \beta(2/\alpha,l-2/\alpha)$, based on \cite{GradshteynRyzhik_2007}[Equation 3.194.3].

Finally, after we plug Eq.~(33) and Eq.~(34) into Eq.~(32), and, subsequently, de-condition on $r$, we get the expression for the coverage probability of operator $n$ as in Eq.~(18).

%BIBLIOGRAPHY
\ifCLASSOPTIONcaptionsoff
  \newpage
\fi

\bibliographystyle{./IEEEtran}
\bibliography{./IEEEabrv,./IEEEfull}

\end{document}